\PassOptionsToPackage{sort&compress}{natbib}
\documentclass[pmlr,twocolumn,10pt]{jmlr} 

\jmlrSuppressPackageChecks
\jmlrvolume{297}
\jmlryear{2025}
\jmlrworkshop{Machine Learning for Health (ML4H) 2025}
\usepackage{booktabs}
\usepackage{siunitx}

\usepackage[utf8]{inputenc}
\usepackage[T1]{fontenc}
\usepackage{microtype}
\usepackage{nicefrac}
\usepackage{amsfonts}
\usepackage{xcolor}
\usepackage{tikz}
\usepackage{graphicx}
\usepackage{overpic}
\usetikzlibrary{arrows.meta,positioning,shapes,spy,calc}
\usepackage{multirow}
\usepackage{subcaption} 
\usepackage{algorithm}   
\usetikzlibrary{spy}
\usepackage{adjustbox}
\usepackage{algorithmic}
\usepackage{tabularx,booktabs,threeparttable,array}
\newcolumntype{L}[1]{>{\raggedright\arraybackslash\hsize=#1\hsize}X}
\newcolumntype{R}[1]{>{\centering\arraybackslash\hsize=#1\hsize}X}

\newif\ifblackandwhitecycle
\gdef\patternnumber{0}

\pgfkeys{/tikz/.cd,
    zoombox paths/.style={
        draw=orange,
        very thick
    },
    black and white/.is choice,
    black and white/.default=static,
    black and white/static/.style={
        draw=white,
        zoombox paths/.append style={
            draw=white,
            postaction={
                draw=black,
                loosely dashed
            }
        }
    },
    black and white/static/.code={
        \gdef\patternnumber{1}
    },
    black and white/cycle/.code={
        \blackandwhitecycletrue
        \gdef\patternnumber{1}
    },
    black and white pattern/.is choice,
    black and white pattern/0/.style={},
    black and white pattern/1/.style={
            draw=white,
            postaction={
                draw=black,
                dash pattern=on 2pt off 2pt
            }
    },
    black and white pattern/2/.style={
            draw=white,
            postaction={
                draw=black,
                dash pattern=on 4pt off 4pt
            }
    },
    black and white pattern/3/.style={
            draw=white,
            postaction={
                draw=black,
                dash pattern=on 4pt off 4pt on 1pt off 4pt
            }
    },
    black and white pattern/4/.style={
            draw=white,
            postaction={
                draw=black,
                dash pattern=on 4pt off 2pt on 2 pt off 2pt on 2 pt off 2pt
            }
    },
    zoomboxarray inner gap/.initial=5pt,
    zoomboxarray columns/.initial=2,
    zoomboxarray rows/.initial=1,
    zoomboxarray heightmultiplier/.initial=0.5,
    subfigurename/.initial={},
    figurename/.initial={zoombox},
    zoomboxarray/.style={
        execute at begin picture={
            \begin{scope}[
                spy using outlines={%
                    zoombox paths,
                    width=\imagewidth / \pgfkeysvalueof{/tikz/zoomboxarray columns} - (\pgfkeysvalueof{/tikz/zoomboxarray columns} - 1) / \pgfkeysvalueof{/tikz/zoomboxarray columns} * \pgfkeysvalueof{/tikz/zoomboxarray inner gap} -\pgflinewidth,
                    height=\pgfkeysvalueof{/tikz/zoomboxarray heightmultiplier} * (\imageheight / \pgfkeysvalueof{/tikz/zoomboxarray rows} - (\pgfkeysvalueof{/tikz/zoomboxarray rows} - 1) / \pgfkeysvalueof{/tikz/zoomboxarray rows} * \pgfkeysvalueof{/tikz/zoomboxarray inner gap}-\pgflinewidth),
                    magnification=3,
                    every spy on node/.style={
                        zoombox paths
                    },
                    every spy in node/.style={
                        zoombox paths
                    }
                }
            ]
        },
        execute at end picture={
            \end{scope}
     \gdef\patternnumber{0}
        },
        spymargin/.initial=0.5em,
        zoomboxes xshift/.initial=1,
        zoomboxes right/.code=\pgfkeys{/tikz/zoomboxes xshift=1},
        zoomboxes left/.code=\pgfkeys{/tikz/zoomboxes xshift=-1},
        zoomboxes yshift/.initial=0,
        zoomboxes above/.code={
            \pgfkeys{/tikz/zoomboxes yshift=1},
            \pgfkeys{/tikz/zoomboxes xshift=0}
        },
        zoomboxes below/.code={
            \pgfkeys{/tikz/zoomboxes yshift=-1},
            \pgfkeys{/tikz/zoomboxes xshift=0}
        },
        caption margin/.initial=0ex, %
    },
    adjust caption spacing/.code={},
    image container/.style={
        inner sep=0pt,
        at=(image.north),
        anchor=north,
        adjust caption spacing
    },
    zoomboxes container/.style={
        inner sep=0pt,
        at=(image.north),
        anchor=north,
        name=zoomboxes container,
        xshift=\pgfkeysvalueof{/tikz/zoomboxes xshift}*(\imagewidth+\pgfkeysvalueof{/tikz/spymargin}),
        yshift=\pgfkeysvalueof{/tikz/zoomboxes yshift}*(\imageheight+\pgfkeysvalueof{/tikz/spymargin}+\pgfkeysvalueof{/tikz/caption margin}),
        adjust caption spacing
    },
    calculate dimensions/.code={
        \pgfpointdiff{\pgfpointanchor{image}{south west} }{\pgfpointanchor{image}{north east} }
        \pgfgetlastxy{\imagewidth}{\imageheight}
        \global\let\imagewidth=\imagewidth
        \global\let\imageheight=\imageheight
        \gdef\columncount{1}
        \gdef\rowcount{1}
        
    },
    image node/.style={
        inner sep=0pt,
        name=image,
        anchor=south west,
        append after command={
            [calculate dimensions]
            node [image container,subfigurename=\pgfkeysvalueof{/tikz/figurename}-image] {\phantomimage}
            node [zoomboxes container,subfigurename=\pgfkeysvalueof{/tikz/figurename}-zoom] {\phantomimage}
        }
    },
    color code/.style={
        zoombox paths/.append style={draw=#1}
    },
    connect zoomboxes/.style={
    spy connection path={\draw[draw=none,zoombox paths] (tikzspyonnode) -- (tikzspyinnode);}
    },
    help grid code/.code={
        \begin{scope}[
                x={(image.south east)},
                y={(image.north west)},
                font=\footnotesize,
                help lines,
                overlay
            ]
            \foreach \x in {0,1,...,9} {
                \draw(\x/10,0) -- (\x/10,1);
                \node [anchor=north] at (\x/10,0) {0.\x};
            }
            \foreach \y in {0,1,...,9} {
                \draw(0,\y/10) -- (1,\y/10);                        \node [anchor=east] at (0,\y/10) {0.\y};
            }
        \end{scope}
    },
    help grid/.style={
        append after command={
            [help grid code]
        }
    },
}

\newcommand\phantomimage{%
    \phantom{%
        \rule{\imagewidth}{\imageheight}%
    }%
}
\newcommand\zoombox[2][]{
    \begin{scope}[zoombox paths]
        \pgfmathsetmacro\xpos{
            (\columncount-1)*(\imagewidth / \pgfkeysvalueof{/tikz/zoomboxarray columns} + \pgfkeysvalueof{/tikz/zoomboxarray inner gap} / \pgfkeysvalueof{/tikz/zoomboxarray columns} ) + \pgflinewidth
        }
        \pgfmathsetmacro\ypos{
            (\rowcount-1) * (\imageheight / \pgfkeysvalueof{/tikz/zoomboxarray rows} + \pgfkeysvalueof{/tikz/zoomboxarray inner gap} / \pgfkeysvalueof{/tikz/zoomboxarray rows} ) + 0.5*\pgflinewidth
        }
        \edef\dospy{\noexpand\spy [
            #1,
            zoombox paths/.append style={
                black and white pattern=\patternnumber
            },
            every spy on node/.append style={#1},
            x=\imagewidth,
            y=\imageheight
        ] on (#2) in node [anchor=north west] at ($(zoomboxes container.north west)+(\xpos pt,-\ypos pt)$);}
        \dospy
        \pgfmathtruncatemacro\pgfmathresult{ifthenelse(\columncount==\pgfkeysvalueof{/tikz/zoomboxarray columns},\rowcount+1,\rowcount)}
        \global\let\rowcount=\pgfmathresult
        \pgfmathtruncatemacro\pgfmathresult{ifthenelse(\columncount==\pgfkeysvalueof{/tikz/zoomboxarray columns},1,\columncount+1)}
        \global\let\columncount=\pgfmathresult
        \ifblackandwhitecycle
            \pgfmathtruncatemacro{\newpatternnumber}{\patternnumber+1}
            \global\edef\patternnumber{\newpatternnumber}
        \fi
    \end{scope}
}

\definecolor{MRIcolA}{HTML}{e0d291}
\definecolor{MRIcolB}{HTML}{d66079}

\title[Patch-Based Diffusion for Data-Efficient MRI Reconstruction]{PaDIS-MRI: Patch-Based Diffusion for Data-Efficient, \\ Radiologist-Preferred MRI Reconstruction}

\author{%
\Name{Rohan Sanda} \Email{rsanda@stanford.edu}\\
\addr Department of Computer Science, Stanford University, USA
\AND
\Name{Asad Aali} \Email{asadaali@stanford.edu}\\
\addr Department of Radiology, Stanford University School of Medicine, USA
\AND
\Name{Andrew Johnston} \Email{drewj32@stanford.edu}\\
\addr Department of Radiology, Stanford University School of Medicine, USA
\AND
\Name{Eduardo Reis} \Email{edreis@stanford.edu}\\
\addr Stanford Center for Artificial Intelligence in Medicine and Imaging, USA 
\AND
\Name{Gordon Wetzstein} \Email{gordonwz@stanford.edu}\\
\addr Department of Electrical Engineering, Stanford University, USA
\AND
\Name{Sara Fridovich-Keil} \Email{sfk@gatech.edu}\\
\addr School of Electrical and Computer Engineering, Georgia Institute of Technology, USA
}

\begin{document}
\maketitle

\begin{abstract}
Magnetic resonance imaging (MRI) requires long acquisition times, which raise costs, reduce accessibility, and increase susceptibility to motion artifacts. Diffusion probabilistic models that learn data-driven priors may reduce acquisition time by enabling reconstruction from undersampled k-space measurements. However, they typically require large training datasets that can be prohibitively expensive to collect. Patch-based diffusion models have shown promise in learning effective data-driven priors over small real-valued datasets, but have not yet demonstrated clinical value in MRI.  We extend the Patch-based Diffusion Inverse Solver (PaDIS) to complex-valued, multi-coil MRI reconstruction, and compare it against a state-of-the-art whole-image diffusion baseline (FastMRI-EDM) for $7\times$ undersampled MRI reconstruction on the FastMRI brain dataset. 
We show that PaDIS-MRI models trained on small datasets of as few as 25 k-space images outperform FastMRI-EDM on image quality metrics (PSNR, SSIM, NRMSE), pixel-level mask-induced variability, cross-contrast/-modality generalization, and robustness to severe k-space undersampling. In a blinded study with three radiologists, PaDIS-MRI reconstructions were chosen as diagnostically superior in $91.7\%$ of cases, compared to baselines (i) FastMRI-EDM and (ii) classical convex reconstruction with wavelet sparsity. These findings highlight the potential of patch-based diffusion priors for high‐fidelity MRI reconstruction in data‐scarce clinical settings where diagnostic confidence matters.
\end{abstract}


\paragraph*{Data and Code Availability.}
Code is available at: \href{https://github.com/voilalab/PaDIS-MRI}{https://github.com/voilalab/PaDIS-MRI}. Data is already public as part of FastMRI \citep{zbontar2019fastmriopendatasetbenchmarks}.

\paragraph*{Institutional Review Board (IRB).}
This work uses the public, de-identified FastMRI dataset \citep{zbontar2019fastmriopendatasetbenchmarks}, with clinical analysis provided by radiologist members of the research team. IRB review was not required.

\section{Introduction}

\begin{figure}[ht]
    \centering
    \includegraphics[width=1.02\linewidth]{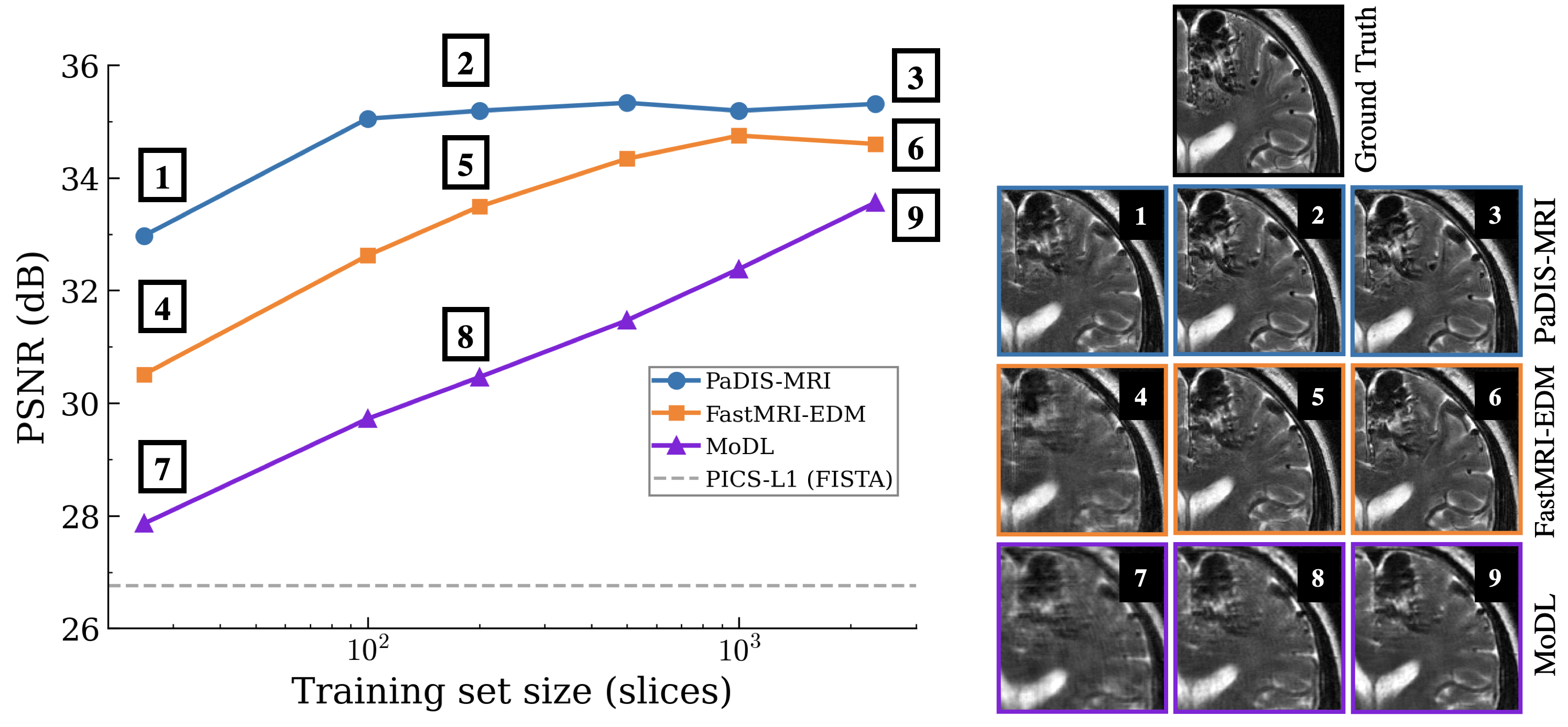}
    \caption{\textbf{Reconstruction quality across dataset sizes.} PaDIS-MRI (ours) consistently outperforms FastMRI-EDM (a full-image diffusion prior) and MoDL (a learned, supervised prior) across all training set sizes, even with as few as 25 images. PICS-L1 is untrained and thus independent of training set size. Tabulated metrics are in Appendix~\ref{appendix:psnr_ds}.}
    \label{fig:psnr_ds}
\end{figure}

Magnetic Resonance Imaging (MRI) is one of the most versatile and non-invasive diagnostic tools in modern medicine. Unlike other popular imaging modalities such as X-ray or computed tomography (CT) scans, MRI leverages magnetic fields and radiofrequency pulses to visualize soft tissue structures with high contrast without cell-damaging ionizing radiation \citep{grover2015magnetic}. Despite its widespread use in clinical practice, the long acquisition times of MRI \citep{hollingsworth2015reducing} contribute to increased cost, patient discomfort, and degraded image quality from motion artifacts \citep{jaspan2015compressed}. 

To decrease acquisition time, it is common practice for modern scanners to undersample k-space – the Fourier domain in which MRI data is captured \citep{ye2019compressed}. However, this introduces an ill-posed inverse problem of reconstructing a high-fidelity image from incomplete measurements. MRI reconstruction algorithms surmount this challenge by leveraging structural priors over medical images, ranging from classical compressed sensing priors like wavelet sparsity and limited total variation to modern data-driven priors that capture the statistics of training images \citep{ye2019compressed}. Among these approaches, generative models (e.g. GANs, VAEs, and diffusion models) have emerged as powerful tools for learning complex, high-dimensional priors directly from MRI data \citep{aali2023solving, aali2024robustmulticoilmrireconstruction, aali2024ambient, fan2024survey}. Through a Bayesian framework, these models allow reconstruction to be posed as a posterior sampling optimization problem, where the goal is to find images that both explain the measured data and conform to the learned prior distribution of the anatomy observed in fully-sampled training images \citep{jalal2021robustcompressedsensingmri}.

\begin{figure*}[!t]
    \centering
    \resizebox{\textwidth}{!}{%
    \begin{tabular}{@{}c@{\hspace{2pt}}c@{\hspace{2pt}}c@{\hspace{2pt}}c@{\hspace{2pt}}c@{\hspace{2pt}}c@{}}
        \begin{tikzpicture}[spy using outlines={circle, magnification=3, size=1.5cm, connect spies}]
            \node[anchor=south west,inner sep=0] (image) at (0,0) {\adjustbox{cfbox=black 0pt,min size={0.155\textwidth}{0.155\textwidth}}{\includegraphics[width=0.155\textwidth]{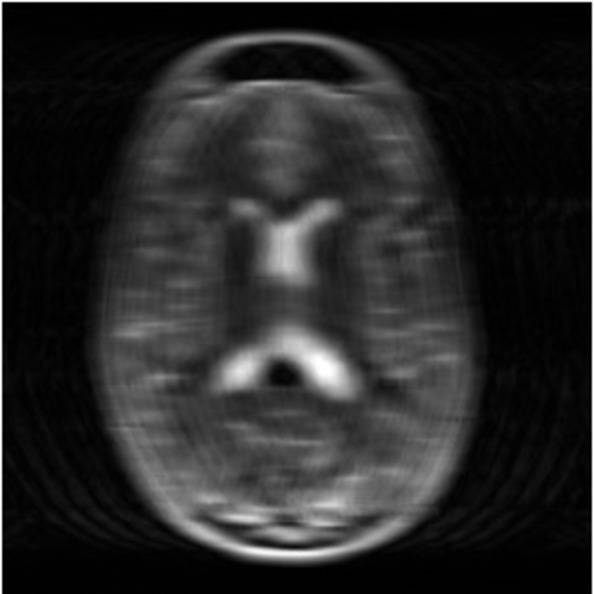}}};
            \spy [red, size=1.2cm] on (1.3,1.5) in node [right] at (image.north west);
        \end{tikzpicture} &
        \begin{tikzpicture}[spy using outlines={circle, magnification=3, size=1.5cm, connect spies}]
            \node[anchor=south west,inner sep=0] (image) at (0,0) {\adjustbox{cfbox=black 0pt,min size={0.155\textwidth}{0.155\textwidth}}{\includegraphics[width=0.155\textwidth]{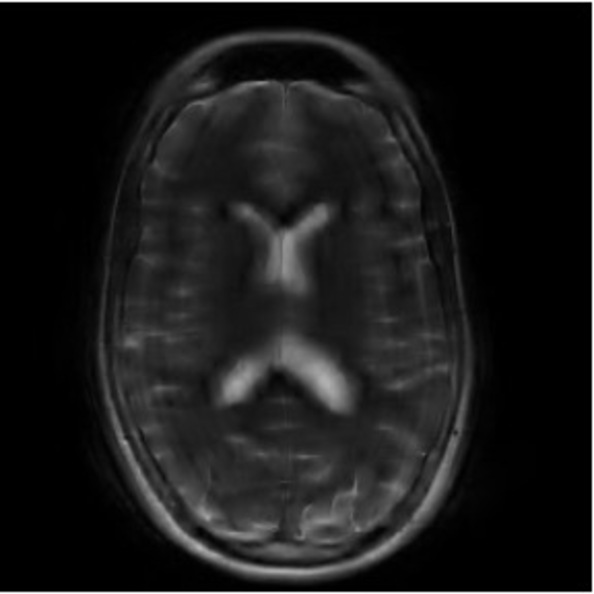}}};
            \spy [red, size=1.2cm] on (1.3,1.5) in node [right] at (image.north west);
        \end{tikzpicture} &
        \begin{tikzpicture}[spy using outlines={circle, magnification=3, size=1.5cm, connect spies}]
            \node[anchor=south west,inner sep=0] (image) at (0,0) {\adjustbox{cfbox=black 0pt,min size={0.155\textwidth}{0.155\textwidth}}{\includegraphics[width=0.155\textwidth]{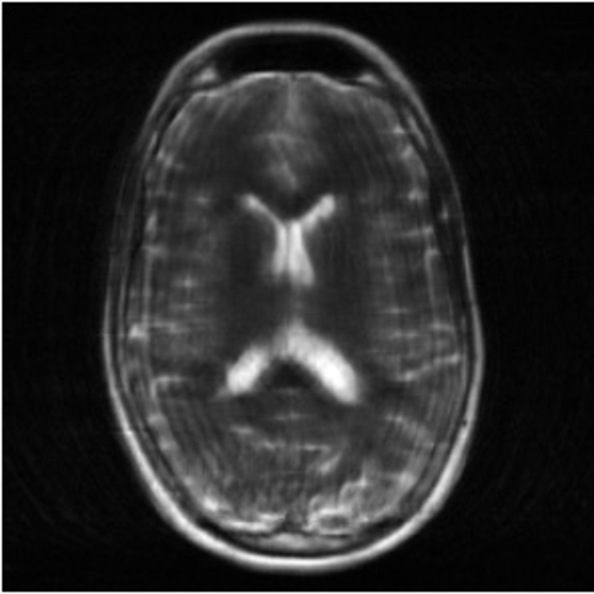}}};
            \spy [red, size=1.2cm] on (1.3,1.5) in node [right] at (image.north west);
        \end{tikzpicture} &
        \begin{tikzpicture}[spy using outlines={circle, magnification=3, size=1.5cm, connect spies}]
            \node[anchor=south west,inner sep=0] (image) at (0,0) {\adjustbox{cfbox=black 0pt,min size={0.155\textwidth}{0.155\textwidth}}{\includegraphics[width=0.155\textwidth]{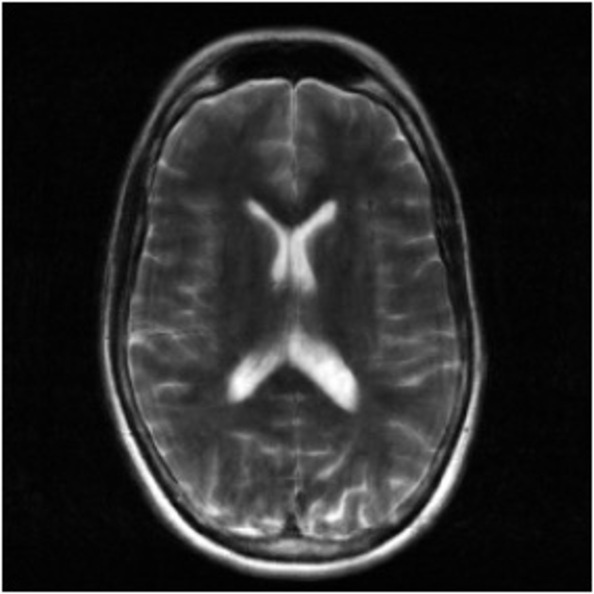}}};
            \spy [red, size=1.2cm] on (1.3,1.5) in node [right] at (image.north west);
        \end{tikzpicture} &
        \begin{tikzpicture}[spy using outlines={circle, magnification=3, size=1.5cm, connect spies}]
            \node[anchor=south west,inner sep=0] (image) at (0,0) {\adjustbox{cfbox=black 0pt,min size={0.155\textwidth}{0.155\textwidth}}{\includegraphics[width=0.155\textwidth]{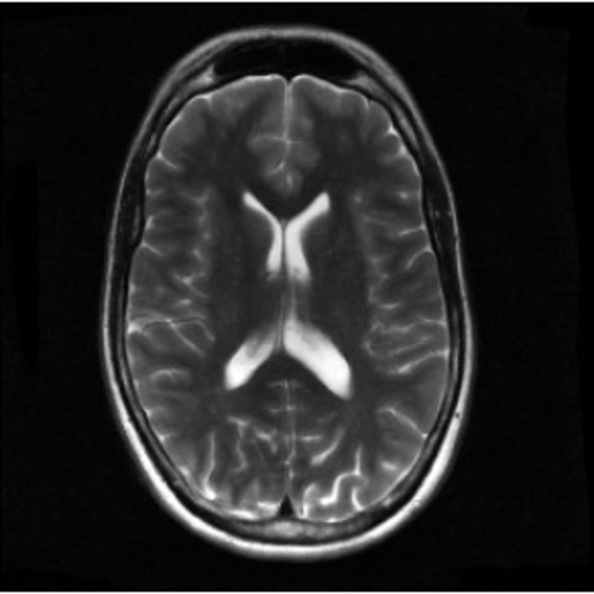}}};
            \spy [red, size=1.2cm] on (1.3,1.5) in node [right] at (image.north west);
        \end{tikzpicture} &
        \begin{tikzpicture}[spy using outlines={circle, magnification=3, size=1.5cm, connect spies}]
            \node[anchor=south west,inner sep=0] (image) at (0,0) {\adjustbox{cfbox=black 0pt,min size={0.155\textwidth}{0.155\textwidth}}{\includegraphics[width=0.155\textwidth]{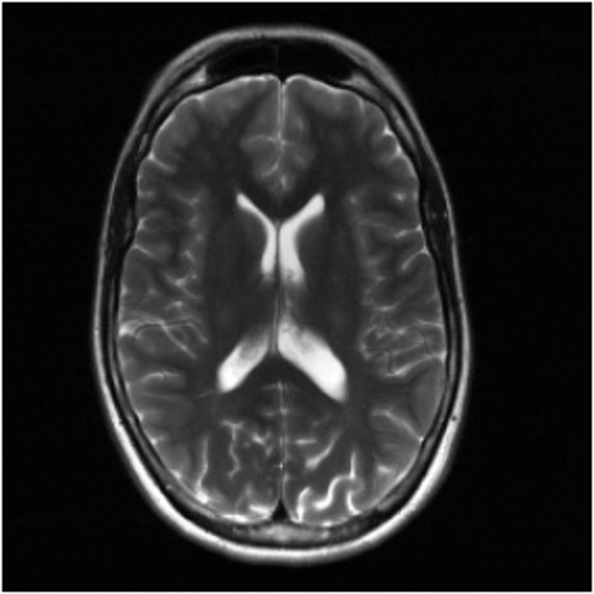}}};
            \spy [red, size=1.2cm] on (1.3,1.5) in node [right] at (image.north west);
        \end{tikzpicture} \\
        
        \begin{tikzpicture}[spy using outlines={circle, magnification=3, size=1.5cm, connect spies}]
            \node[anchor=south west,inner sep=0] (image) at (0,0) {\adjustbox{cfbox=black 0pt,min size={0.155\textwidth}{0.155\textwidth}}{\includegraphics[width=0.155\textwidth]{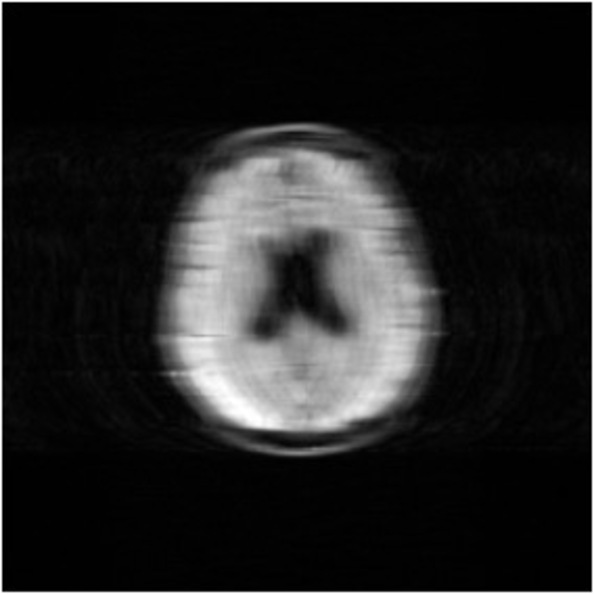}}};
            \spy [red, size=1.2cm] on (1.28,1.1) in node [right] at (image.north west);
        \end{tikzpicture} &
        \begin{tikzpicture}[spy using outlines={circle, magnification=3, size=1.5cm, connect spies}]
            \node[anchor=south west,inner sep=0] (image) at (0,0) {\adjustbox{cfbox=black 0pt,min size={0.155\textwidth}{0.155\textwidth}}{\includegraphics[width=0.155\textwidth]{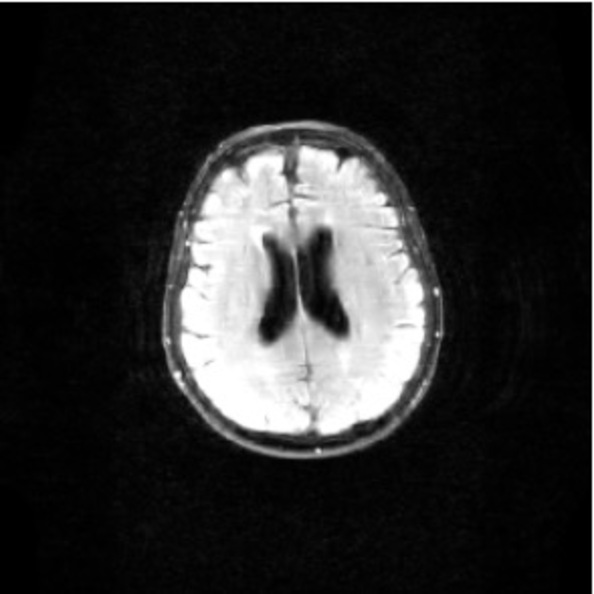}}};
            \spy [red, size=1.2cm] on (1.3,1.1) in node [right] at (image.north west);
        \end{tikzpicture} &
        \begin{tikzpicture}[spy using outlines={circle, magnification=3, size=1.5cm, connect spies}]
            \node[anchor=south west,inner sep=0] (image) at (0,0) {\adjustbox{cfbox=black 0pt,min size={0.155\textwidth}{0.155\textwidth}}{\includegraphics[width=0.155\textwidth]{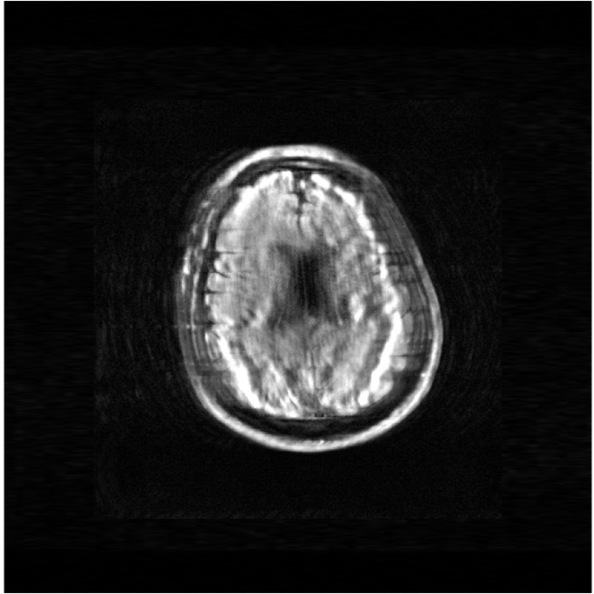}}};
            \spy [red, size=1.2cm] on (1.3,1.14) in node [right] at (image.north west);
        \end{tikzpicture} &
        \begin{tikzpicture}[spy using outlines={circle, magnification=3, size=1.5cm, connect spies}]
            \node[anchor=south west,inner sep=0] (image) at (0,0) {\adjustbox{cfbox=black 0pt,min size={0.155\textwidth}{0.155\textwidth}}{\includegraphics[width=0.155\textwidth]{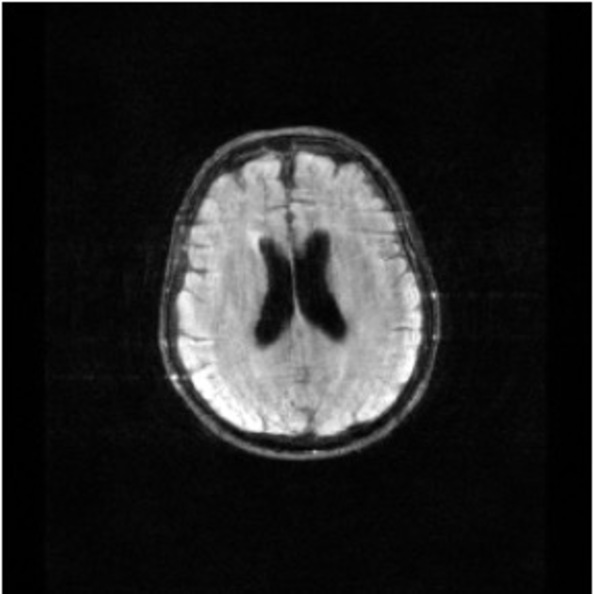}}};
            \spy [red, size=1.2cm] on (1.3,1.1) in node [right] at (image.north west);
        \end{tikzpicture} &
        \begin{tikzpicture}[spy using outlines={circle, magnification=3, size=1.5cm, connect spies}]
            \node[anchor=south west,inner sep=0] (image) at (0,0) {\adjustbox{cfbox=black 0pt,min size={0.155\textwidth}{0.155\textwidth}}{\includegraphics[width=0.155\textwidth]{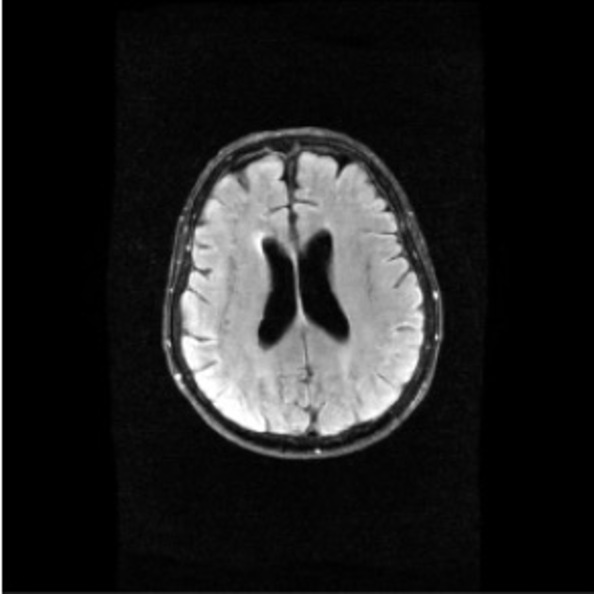}}};
            \spy [red, size=1.2cm] on (1.3,1.1) in node [right] at (image.north west);
        \end{tikzpicture} &
        \begin{tikzpicture}[spy using outlines={circle, magnification=3, size=1.5cm, connect spies}]
            \node[anchor=south west,inner sep=0] (image) at (0,0) {\adjustbox{cfbox=black 0pt,min size={0.155\textwidth}{0.155\textwidth}}{\includegraphics[width=0.155\textwidth]{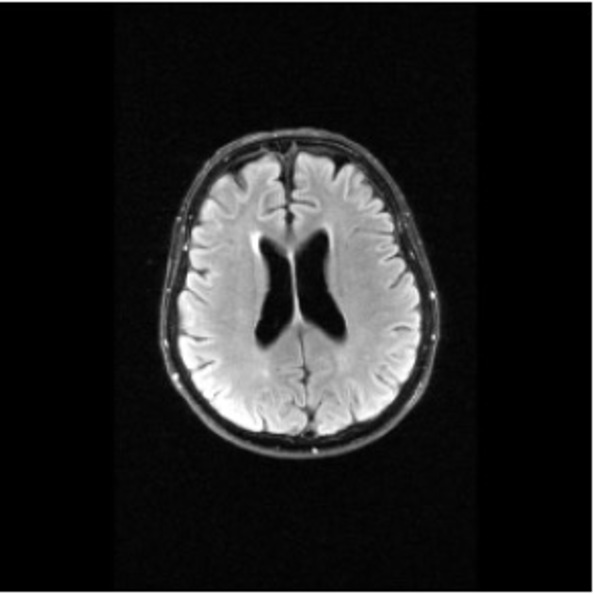}}};
            \spy [red, size=1.2cm] on (1.3,1.1) in node [right] at (image.north west);
        \end{tikzpicture} \\
        
        \begin{tikzpicture}[spy using outlines={circle, magnification=3, size=1.5cm, connect spies}]
            \node[anchor=south west,inner sep=0] (image) at (0,0) {\adjustbox{cfbox=black 0pt,min size={0.155\textwidth}{0.155\textwidth}}{\includegraphics[width=0.155\textwidth]{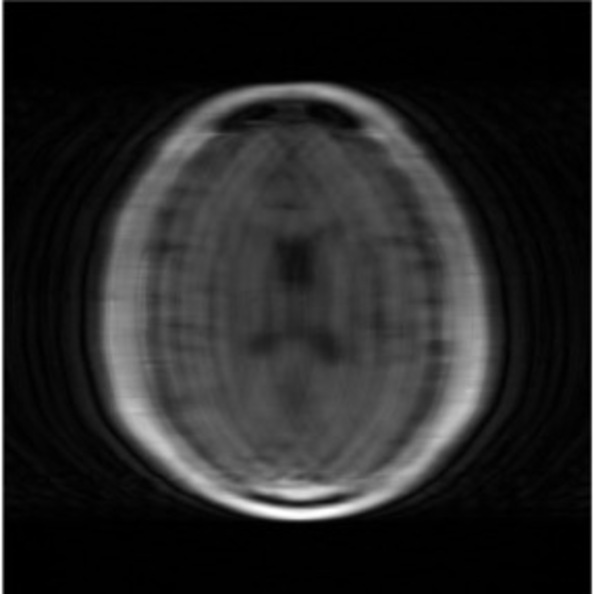}}};
            \spy [red, size=1.2cm] on (1.48,0.8) in node [right] at (image.north west);
        \end{tikzpicture} &
        \begin{tikzpicture}[spy using outlines={circle, magnification=3, size=1.5cm, connect spies}]
            \node[anchor=south west,inner sep=0] (image) at (0,0) {\adjustbox{cfbox=black 0pt,min size={0.155\textwidth}{0.155\textwidth}}{\includegraphics[width=0.155\textwidth]{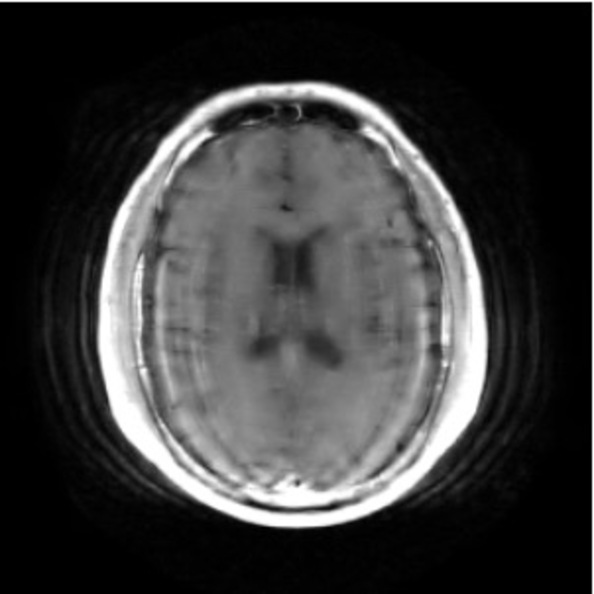}}};
            \spy [red, size=1.2cm] on (1.48,0.8) in node [right] at (image.north west);
        \end{tikzpicture} &
        \begin{tikzpicture}[spy using outlines={circle, magnification=3, size=1.5cm, connect spies}]
            \node[anchor=south west,inner sep=0] (image) at (0,0) {\adjustbox{cfbox=black 0pt,min size={0.155\textwidth}{0.155\textwidth}}{\includegraphics[width=0.155\textwidth]{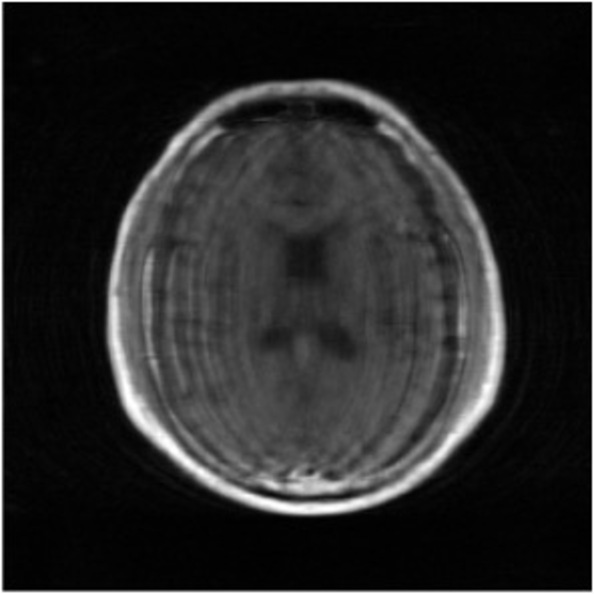}}};
            \spy [red, size=1.2cm] on (1.48,0.8) in node [right] at (image.north west);
        \end{tikzpicture} &
        \begin{tikzpicture}[spy using outlines={circle, magnification=3, size=1.5cm, connect spies}]
            \node[anchor=south west,inner sep=0] (image) at (0,0) {\adjustbox{cfbox=black 0pt,min size={0.155\textwidth}{0.155\textwidth}}{\includegraphics[width=0.155\textwidth]{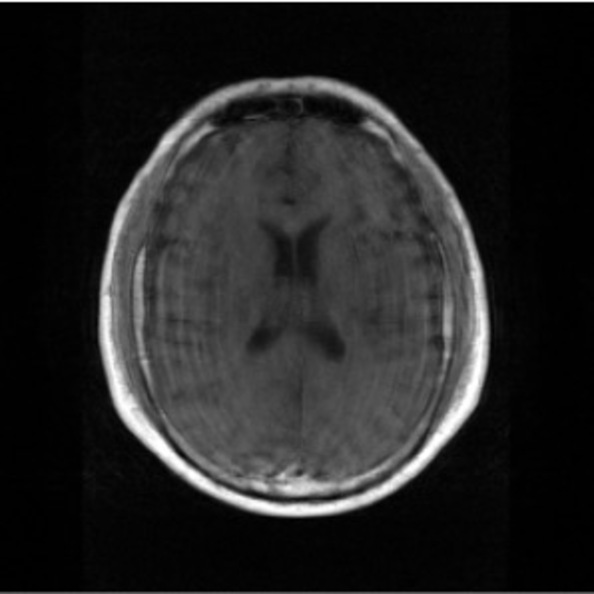}}};
            \spy [red, size=1.2cm] on (1.48,0.8) in node [right] at (image.north west);
        \end{tikzpicture} &
        \begin{tikzpicture}[spy using outlines={circle, magnification=3, size=1.5cm, connect spies}]
            \node[anchor=south west,inner sep=0] (image) at (0,0) {\adjustbox{cfbox=black 0pt,min size={0.155\textwidth}{0.155\textwidth}}{\includegraphics[width=0.155\textwidth]{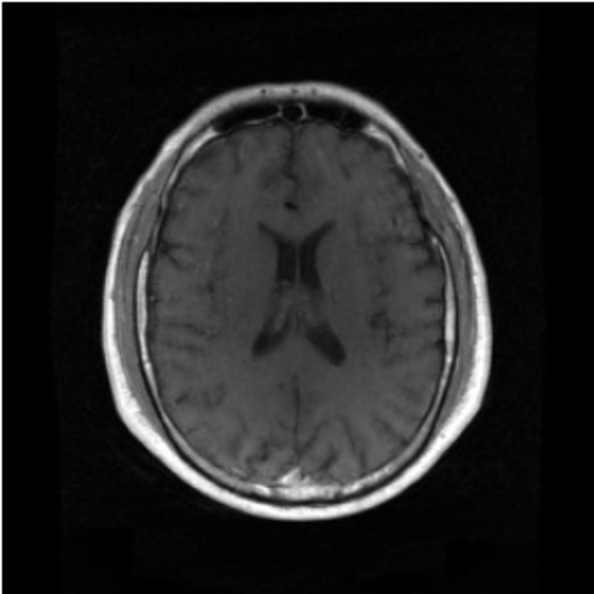}}};
            \spy [red, size=1.2cm] on (1.48,0.8) in node [right] at (image.north west);
        \end{tikzpicture} &
        \begin{tikzpicture}[spy using outlines={circle, magnification=3, size=1.5cm, connect spies}]
            \node[anchor=south west,inner sep=0] (image) at (0,0) {\adjustbox{cfbox=black 0pt,min size={0.155\textwidth}{0.155\textwidth}}{\includegraphics[width=0.155\textwidth]{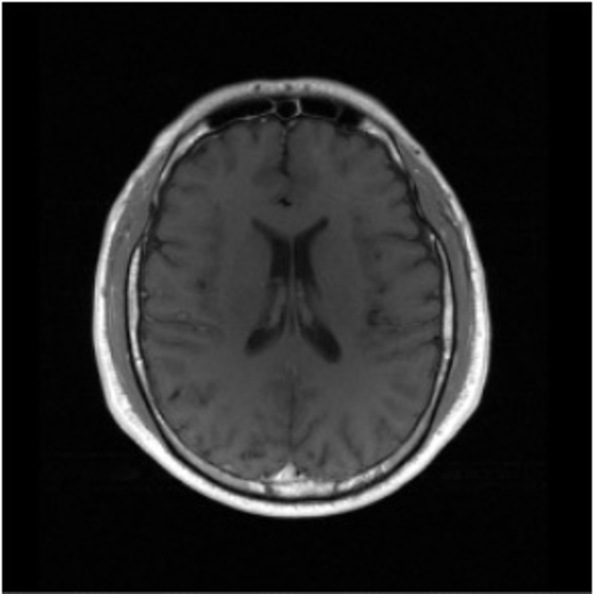}}};
            \spy [red, size=1.2cm] on (1.48,0.8) in node [right] at (image.north west);
        \end{tikzpicture} \\[2pt]
        
        \footnotesize Adjoint & 
        \footnotesize PICS-L1 & 
        \footnotesize MoDL & 
        \footnotesize FastMRI-EDM & 
        \footnotesize \textbf{PaDIS-MRI (ours)} & 
        \footnotesize Ground Truth
    \end{tabular}
    }
    \caption{Example MRI magnitude reconstructions with $S = 25$ training slices at acceleration factor $R = 7$. PaDIS-MRI reconstructions are sharper and more faithful to the fully-sampled ground truth. Insets are shown to highlight detail.}
    \label{fig:recons_scaled}
\end{figure*}

However, generative models require notoriously large training datasets for reliable performance. This dependence presents multiple challenges to clinical use. First, acquiring high-fidelity MRI scans from a diverse set of patients is expensive, time-intensive, and often restricted by privacy constraints. Second, generative models trained on one anatomy (e.g., brain), pathology (e.g., tumor), or contrast (e.g., T1-weighted) may not generalize well to others, limiting real-world deployment. Data-efficient generative priors are thus critical to safe and clinically effective adoption of diffusion priors for MRI reconstruction.

To address these limitations, we extend the Patch-based Diffusion Inverse Solver (PaDIS) framework proposed by \citet{hu2024learningimagepriorspatchbased} to the setting of multi-coil undersampled MRI reconstruction. Rather than learning global anatomical priors over entire images, we train a diffusion model on complex-valued image patches drawn at random from the FastMRI dataset \citep{zbontar2019fastmriopendatasetbenchmarks}. This patch-based formulation enables the model to learn localized structural motifs, improving performance in low-data regimes commonly encountered in clinical settings. 

We integrate this learned patch-based prior into the Diffusion Posterior Sampling (DPS) algorithm to reconstruct MRI images from undersampled k-space measurements, with our modifications specifically designed to handle complex-valued MRI. We benchmark our patch-based PaDIS-MRI prior against the whole-image diffusion model FastMRI-EDM developed by~\citet{aali2024robustmulticoilmrireconstruction}. We find that the patch-based model consistently outperforms the whole-image prior across reconstruction quality metrics (PSNR, SSIM, and NRMSE) using small training datasets with as few as 25 images (see Figure~\ref{fig:psnr_ds}), while also achieving better cross-contrast/-domain generalization and more consistent results across varying undersampling patterns. 

To assess clinical utility, we conducted a blinded comparison analysis in which three radiologists examined 60 sets of images reconstructed by PaDIS-MRI, FastMRI-EDM, and a classical wavelet-sparse reconstruction and selected which image was most faithful to the ground truth fully-sampled reconstruction in terms of diagnostic quality. The majority of radiologists selected PaDIS-MRI as the most clinically accurate reconstruction in 55/60 cases (91.7\%). To summarize, our core contributions include:
\begin{itemize}
    \item Adapting PaDIS \citep{hu2024patchbaseddiffusionmodelsbeat} for training on two-channel (real/imaginary) complex-valued data in undersampled MRI reconstruction.
    \item Evaluating the quantitative reconstruction fidelity of PaDIS-MRI, which outperforms FastMRI-EDM and a wavelet-sparse baseline across training set sizes and k-space undersampling levels, and exhibits better cross-contrast/-domain generalization and lower pixel-wise mask-induced variability.
    \item Evaluating the potential diagnostic utility of PaDIS-MRI compared to FastMRI-EDM and a wavelet-sparse baseline via a blinded comparison study with three radiologists, who overwhelmingly selected PaDIS-MRI reconstructions as having higher diagnostic quality and clinical accuracy with respect to the fully-sampled ground truth.
\end{itemize}

\section{Related Work}
\label{related}

\subsection{Diffusion Models for MRI Reconstruction}

Diffusion models have emerged as powerful generative priors for solving ill-posed inverse problems in MRI reconstruction. \citet{song2021scorebasedgenerativemodelingstochastic} first introduced continuous-time score-based generative modeling, where image generation is cast as reversing a stochastic differential equation (SDE) from noise back to data. Building on this foundation, \citet{jalal2021robustcompressedsensingmri} were among the first to apply score-based models to compressed-sensing MRI, showing that posterior sampling with a learned score prior can reconstruct high fidelity images from undersampled k-space. \citet{chung2022scorebaseddiffusionmodelsaccelerated} adapted score-based diffusion to MRI denoising tasks and demonstrated significant gains over supervised baselines. \citet{chung2024diffusionposteriorsamplinggeneral} then introduced the diffusion posterior sampling (DPS) algorithm by augmenting the reverse SDE with a measurement-consistency gradient to more directly approximate the intractable posterior distribution.

\subsection{Patch-Based Diffusion}

\citet{wang2023patchdiffusionfasterdataefficient} proposed training on image patches to improve efficiency, but still required full-image score estimation at inference time. \citet{hu2024learningimagepriorspatchbased} addressed this limitation with their PaDIS framework, which was demonstrated on a variety of inverse problems involving CT scans and natural images. By modeling an image as a Markov Random Field, where cliques correspond to patches of pixels, they leverage the Hammersley-Clifford theorem to express the score function of the entire image as a sum of score functions of patches. 
More recent work by \citet{hu2024patchbaseddiffusionmodelsbeat} demonstrated that PaDIS also improves out-of-distribution (OOD) generalization compared to whole-image diffusion models. 

A few studies have demonstrated the feasibility of patch-based models for MRI tasks. \citet{behrendt2023patcheddiffusionmodelsunsupervised} introduced patched denoising diffusion probabilistic models for unsupervised anomaly detection in brain MRIs, demonstrating improvements in tumor segmentation performance. Separately, \citet{huang_patchmri} developed the PADMr framework that leverages patch-based inputs and a non-Markovian diffusion process to accelerate inference times for MRI reconstruction. Most recently, \citet{roy2025investigatingpatchbasedmri} studied patch-based inference for generalized diffusion priors in MRI reconstruction, examining memory efficiency and performance on denoising and super-resolution tasks. Rather than focusing on computational benefits, we evaluate the potential clinical impact of the PaDIS framework in multi-coil undersampled MRI reconstruction, focusing on data efficiency, robustness (to contrast variation and degree of undersampling), and diagnostic fidelity as assessed by quantitative image metrics and radiologists.

\section{Background}
\label{background}

\subsection{MRI Reconstruction}
\label{background:prob}

Multi-coil MRI is a linear inverse problem that aims to reconstruct a vectorized 2D anatomical ground-truth image \( x \in \mathbb{C}^{N_x N_y} \) from undersampled k-space measurements \( y \). Specifically, the forward operator \( \mathcal{A} \) combines the coil sensitivity operator \( S \in \mathbb{C}^{N_c N_x N_y \times N_x N_y} \), the Fourier Transform operator \( F \in \mathbb{C}^{N_c N_x N_y \times N_c N_x N_y} \), and the undersampling pattern \( P \in \mathbb{C}^{\frac{N_c N_x N_y}{R} \times N_c N_x N_y} \), where \( N_c \) is the number of coils, \( N_x, N_y \) denote spatial dimensions of the image, and \( R \) represents the undersampling (acceleration) factor \citep{aali2024robustmulticoilmrireconstruction}. The resulting undersampled measurements are thus \( y \in \mathbb{C}^{\frac{N_c N_x N_y}{R}} \) and we have:
\begin{equation}
y = P F S x + \epsilon
\end{equation}
where \(\epsilon\) denotes complex-valued additive noise from scanner imperfections or patient motion. Due to k-space undersampling (\(R > 1\)), reconstructing the ground truth clean image \(x\) from \(y\) constitutes an ill-posed inverse problem. Traditionally, this problem is formulated as a regularized optimization over \(x\) where \( \mathcal{A} = PFS\):
\begin{equation}
\tilde{x} = \arg\min_x \| y - \mathcal{A} x \|_2^2 + \lambda \mathcal{R}(x) .
\end{equation}
From a Bayesian perspective, optimization problems like Equation (2) are often formulated as maximum a posteriori (MAP) estimation. However, DPS instead approximates posterior sampling from \(p(x|y) \propto p(y|x)p(x)\) using a learned score-based prior $p_\theta(x)$, yielding diverse generations that reflect the full posterior distribution rather than collapsing to a single, deterministic point estimate.

\subsection{Diffusion and DPS}

 Score-based diffusion models follow an SDE to progressively corrupt a clean image \(x\). The forward SDE, defined by \citet{song2021scorebasedgenerativemodelingstochastic}, can be expressed as:
 \begin{equation}
     dx = f(x,t)\,dt + g(t)\,dW_t,
 \end{equation}
 where \(W_t\) denotes standard Brownian motion, and functions \(f\) and \(g\) define drift and diffusion coefficients, respectively chosen so that at time \(t=T\), \(x\sim\mathcal{N}(0,\sigma_T^2I)\).  Sampling from this model involves solving the reverse-time SDE~\citep{song2021scorebasedgenerativemodelingstochastic}:
 \begin{equation}
     dx = \left[f(x,t) - g(t)^2 \nabla_x \log p_t(x)\right]\,dt + g(t)\,d\bar{W_t},
 \end{equation}
 where \(\nabla_x\log p_t(x)\) is the learned score function estimated by the denoising neural network \(D_\theta\).  
 
 To solve linear inverse problems of the form \( y = \mathcal{A}x + \epsilon\), \citet{chung2024diffusionposteriorsamplinggeneral} augment the reverse SDE with a measurement-consistency likelihood gradient:
\begin{multline}
dx = \bigl[f(x,t) - g(t)^2 \nabla_{\!x}\log p_t(x)\bigr]\,dt\\
\qquad {}+ t\,\nabla_{\!x}\log p(y\mid x)\,dt
        + g(t)\,d\bar W_t .
\end{multline}
 where \(\nabla_x\log p(y\mid x) = -\nabla_x\|y - \mathcal A x\|^2\). In practice, to solve Equation (5) the Variance-Exploding (VE) discretization from \citet{karras2022elucidatingdesignspacediffusionbased} can be used, parameterized by:

\begin{equation}
  t_k 
  = \Bigl(\sigma_{\max}^{1/\rho} 
       + \tfrac{k}{N-1}\bigl(\sigma_{\min}^{1/\rho}-\sigma_{\max}^{1/\rho}\bigr)\Bigr)^{\!\rho},
  \quad k=0,\dots,K-1
\end{equation}
where \(K\) is the number of steps, and \(\rho\), \(\sigma_{\min}\), and \(\sigma_{\max}\) are hyperparameters for the noising schedule. In the VE-DPS algorithm, the following updates to the image \(x_k\) at step \(k\) are made:
\begin{align*}
&\alpha_k = \tfrac12\,t_k^{2}, ~~~ z_k\sim\mathcal N(0,I), ~~~\mathrm{SSE}_k = \bigl\|y - \mathcal A\,\hat x_k\bigr\|_2^2, \\
&\text{score}_k = \frac{D_\theta(x_k,t_k)-x_k}{t_k^{2}}, 
~~\eta_k =
\begin{cases}
\sqrt{\alpha_k}\,z_k, & k<K,\\
0, & k=K,
\end{cases} \\
&x_k' = x_k - \frac{\zeta}{\sqrt{\mathrm{SSE}_k}}\,
        \nabla_{\!x_k}\,\mathrm{SSE}_k, \\
&x_{k+1} = x_k' + \frac{\alpha_k}{2}\,\text{score}_k + \eta_k, 
\end{align*}
where $\zeta$ controls the trade-off between data fidelity and prior consistency and \(y\) and \(\mathcal{A}\) are as defined in Subsection \ref{background:prob}.

\subsection{Diffusion with Patches}
\label{section:background:padis}
The PaDIS framework \citep{hu2024learningimagepriorspatchbased} employs a single neural network to learn a unified prior across all image patches, incorporating positional encodings to distinguish between patch locations within the image. During training, PaDIS adopts a flexible patching strategy where patches are extracted from zero-padded images, with both patch size and location randomized. Each patch is processed alongside its positional encoding, allowing the network to learn location-aware representations across multiple scales. At inference, we construct a fixed coverage grid and randomly choose a start offset at each diffusion iteration. Given an input image $x_0 \in \mathbb{R}^{N \times N}$ and a designated patch size $P$, we zero-pad the image to admit a $(k{+}1)\times(k{+}1)$ grid of $P\times P$ patches with stride $P$, where $k=\lfloor N/P \rfloor$. The padding width $M=(k{+}1)P-N$ ensures that the grid extends beyond the original image boundaries; it also equals the number of valid start offsets per axis, so there are $M^2$ possible partitionings.

The global score function of the entire image is then expressed as in \citet{hu2024learningimagepriorspatchbased}:
\begin{equation*}
    \nabla_x \log p(x)
    = \sum_{i,j=1}^{M} \Big( s_{i,j,B}(x_{i,j,B}) + \sum_{r=1}^{(k+1)^2} s_{i,j,r}(x_{i,j,r}) \Big) ,
\end{equation*}
where each pair $(i,j)$ indexes a start-offset partition, $x_{i,j,r}$ denotes the $r$-th patch under partition $(i,j)$, $s_{i,j,r}$ is its corresponding score function, and $x_{i,j,B}$ is the zero-padded border region. For efficiency, at each inference iteration PaDIS samples a single partition $(i,j)$ rather than evaluating all $M^2$; across iterations this Monte-Carlo averaging over start offsets mitigates patch-boundary artifacts.

\vspace{-0.6cm}

\section{Methods}
\label{methods}

We adapt the PaDIS framework \citep{hu2024learningimagepriorspatchbased} for multi-coil undersampled MRI reconstruction, including both training of the patch-based score model and inference via our DPS sampler. Similar to PaDIS, we build on top of the EDM model architecture from \citet{karras2022elucidatingdesignspacediffusionbased}. Implementation details and hyperparameters may be found in Appendix \ref{sec:details}.

\subsection{PaDIS-MRI Training}
Training images of dimension \(N \times N\) are zero-padded by \(\tfrac{1}{4}N\) pixels on all sides (96 px when \(N{=}384\)). Rather than RGB or single‐channel images, we train on complex coil‐combined MR images. Each extracted \(P\times P\) patch (\(P\!\in\!\{16,32,64\}\), sampled with probabilities \(\{0.2,0.3,0.5\}\)) is represented as a 2–channel real tensor for the real/imag parts and processed by a Song-UNet-based denoiser (\(\sim\)55M parameters). For each patch, we append a 2D positional encoding of its \((x,y)\) location (normalized to \([-1,1]\)), informing the denoiser of the absolute position within the full image. This position-aware training enables the model to learn location-specific features while maintaining coherent full-image reconstructions. As in PaDIS, the score-matching loss is computed on patches sampled from random locations across the padded image, allowing the model to learn the distribution of patches at all positions \citep{hu2024learningimagepriorspatchbased}.

\subsection{PaDIS-MRI Inference}
\label{sec:methods:dps}

At inference time, we reconstruct image \(x\) from undersampled k-space \(y\) using our patch‐based VE-DPS sampler with \(K{=}104\) steps.  We warm start with the adjoint \(x_{0} = \mathcal A^{\dagger}(y)\), then symmetrically zero-padded by \(M{=}64\) pixels per side. We can then follow the VE-DPS flow (Algorithm \ref{alg:dps2} in Appendix \ref{appendix:algo}) with patch-size \(P{=}64\) to solve the MRI reconstruction problem. 

For each step \(k=0,\dots,K-1\), we (1) inject noise \(x\gets x + t_k\varepsilon\) with \(\varepsilon\!\sim\!\mathcal N(0,I)\), where \(x \in \mathbb{C}^{(N+2M) \times (N+2M)}\) is the padded image; (2) extract a non‐overlapping grid of \(P\times P\) patches using a random starting offset, (3) convert each complex patch \(p \in\mathbb C^{P \times P}\) into a 2‐channel real tensor \( p_{\text{real}}\in \mathbb{R}^{P \times P \times 2}\), (4) denoise via \(D_\theta\), and reassemble the denoised patches into \(D \in\mathbb C^{(N+2M) \times (N+2M)}\); (5) form the estimated \(\mathrm{score}=(D - x)/t_k^2\); (6) crop \(D\) to the original field of view to obtain \(\hat{x} \in \mathbb{C}^{N \times N}\), (7) compute the residual \(r=y-\mathcal A(\hat x)\) and \(\mathrm{SSE}=\|r\|^2\) in k-space, and (8) backpropagate and pad to get \(\nabla_x\mathrm{SSE} \in \mathbb{C}^{(N+2M) \times (N+2M)}\), and (9) perform the update 
\[
x \;\gets\; \left( x + \frac{\alpha_k}{2}\,\mathrm{score} + \sqrt{\alpha_k}\,\varepsilon \right) - \frac{\zeta}{\sqrt{\mathrm{SSE}}}\,\nabla_x\,\mathrm{SSE}.
\]
After \(K\) iterations (with \(\zeta{=}3.0\)) we remove the padding to obtain the final reconstruction \(x_{K}\in\mathbb C^{N\times N}\).

\subsection{Dataset}
We use fully-sampled, multicoil k-space data from the NYU Langone Health FastMRI dataset \citep{zbontar2019fastmriopendatasetbenchmarks}, including FLAIR, native T1, pre-contrast T1, post-contrast T1, and axial T2-weighted brain scans \citep{zbontar2019fastmriopendatasetbenchmarks}. In total, the training set contains 446 k-space volumes, consisting of 40 FLAIR, 136 T1-weighted (91 post-contrast, 26 pre-contrast, 19 native), and 270 T2-axial. Our validation set contains 266 volumes (distinct from the training volumes) from which we sample an evaluation test set of center-slices to mirror the training contrast distribution, namely 50 T2-axial, 7 FLAIR, and 25 T1-weighted slices. For our cross-modality generalization experiment in Section ~\ref{cross-contrast}, we evaluated on 30 randomly-selected, fat-suppressed (FS) knee MRI volumes (sampling the 21st slice) from the FastMRI dataset. All metrics are reported over this evaluation test set. Further preprocessing details are in Appendix~\ref{sec:details}.

\subsection{Blinded Radiologist Preference Study}
\label{methods:reader}

We ran a blinded 3-way forced-choice preference study with three radiologists. For each of 60 cases (24 T1/FLAIR, 36 T2), each radiologist was shown the labeled ground truth and three reconstructions – PaDIS-MRI (ours, patch-based diffusion prior), FastMRI-EDM (whole-image diffusion prior from \citet{aali2024robustmulticoilmrireconstruction}), and PICS-L1 (wavelet sparsity prior from \citet{bart}) – with method identities and ordering fully anonymized and randomized per case. The 60 cases were randomly split into two groups: in 30 cases the learned priors (PaDIS-MRI and FastMRI-EDM) were trained on 25 slices, and in 30 cases they were trained on 500 slices; each group preserved the same contrast mix (18 T2, 12 T1/FLAIR). Each radiologist selected a single ``best” reconstruction per case, targeting maximal diagnostic quality.
Each image therefore received three independent votes (one per radiologist). We report (i) per-radiologist preference rates for PaDIS-MRI, and (ii) an image-level majority outcome indicating whether $\ge$2 of 3 radiologists chose PaDIS-MRI for that case. We summarize radiologist preference over all 60 images and separately by contrast type (T1/FLAIR vs.\ T2). We report raw preference proportions as well as Wilson (binomial) 95\% confidence intervals (CIs).

\section{Experiments}
\label{results}
We evaluate PaDIS-MRI and FastMRI-EDM \citep{aali2024robustmulticoilmrireconstruction} across several dimensions, against a convex untrained wavelet-sparse reconstruction PICS-L1 \citep{bart}. While PICS-L1 does not require any training data, its classical prior struggles to reconstruct fine details and sharp edges at the high undersampling rates we study ($R=7$), and is outperformed by both diffusion models. We also include model-based deep learning (MoDL) as a supervised baseline \citep{modl} in our data-efficiency experiments; see Appendix~\ref{appendix:modl} for our SENSE + CG unrolling and training setup. 

We first assess quantitative reconstruction quality as a function of training set size, to validate the data efficiency of the patch-based PaDIS-MRI prior. We then examine reconstruction consistency by visualizing pixel-wise standard deviations across reconstructions from multiple k-space masks of the same slice. Finally, we test robustness to varying k-space undersampling levels and generalization to different contrast types. Across all quantitative experiments, we report Peak Signal-to-Noise Ratio (PSNR), Structural Similarity Index (SSIM), and Normalized Root Mean Square Error (NRMSE) compared to fully-sampled ground truth. Since there is large variation in metrics among individual images, instead of reporting dataset-wide standard deviation in metrics for each method, we report the mean and standard deviation of per-image pairwise differences between PaDIS-MRI and each baseline. These pairwise difference results show that PaDIS-MRI often provides consistent improvement across individual images. We evaluate diagnostic quality via expert visual comparison by a panel of three radiologists, who independently assessed each reconstruction in a blinded comparison study with access to ground truth images, but with algorithms anonymized and presented in random order.

\subsection{Data Efficiency}
We evaluate the performance of PaDIS-MRI, FastMRI-EDM, and MoDL when trained on dataset sizes \(S = \{25, 100, 200, 500, 1000, 2330\} \) k-space slices and tested on our evaluation dataset with a fixed acceleration (undersampling) factor of \(R=7\). Figure \ref{fig:psnr_ds} and Table \ref{tab:combined_results} in Appendix \ref{appendix:psnr_ds} summarize our findings. Across training set sizes 25 through 500, PaDIS-MRI consistently outperforms the baselines on all metrics. At larger dataset sizes ($S \geq 1000$), PaDIS-MRI maintains superior PSNR and NRMSE, while FastMRI-EDM shows marginal improvement in SSIM. This small SSIM advantage is likely caused by the whole-image model's ability to capture global perceptual features when provided sufficient training examples, while PaDIS-MRI's patch-based approach continues to excel at preserving fine details and minimizing overall error (as reflected in better PSNR and NRMSE). PaDIS-MRI exhibits the most significant advantages as dataset size decreases, highlighting the improved data efficiency of the patch-based approach. Importantly, FastMRI-EDM (and PaDIS-MRI) outperforms MoDL, supporting the results in \citet{aali2024robustmulticoilmrireconstruction}. Full tabular results including performance stratified by contrast type are presented in Appendix \ref{appendix:psnr_ds}.

Figure \ref{fig:recons_scaled} provides a visual comparison of the reconstruction quality of each method on example slices. In reconstructions from the $S=25, R=7$ models, FastMRI-EDM exhibits slightly more blurring and loss of fine structural details, particularly in regions with complex anatomical features. PaDIS-MRI, in contrast, preserves sharper boundaries and finer details. This enhanced data efficiency can be attributed to the patch-based prior's ability to learn localized structural motifs more effectively from limited examples. 
Figure \ref{fig:mri_fullpage} in the Appendix shows similar reconstructions at \( S = 200, R=7\) where FastMRI-EDM more closely approaches the performance of PaDIS-MRI but still struggles with highly detailed anatomy. At both dataset sizes, MoDL underperforms the diffusion priors. In Figure~\ref{fig:recons_scaled}, MoDL’s FLAIR reconstruction exhibits pronounced artifacts, likely reflecting reduced generalizability of a supervised prior when the training distribution underrepresents FLAIR.

\begin{figure}[h]
    \centering
    \includegraphics[width=1.0\linewidth]{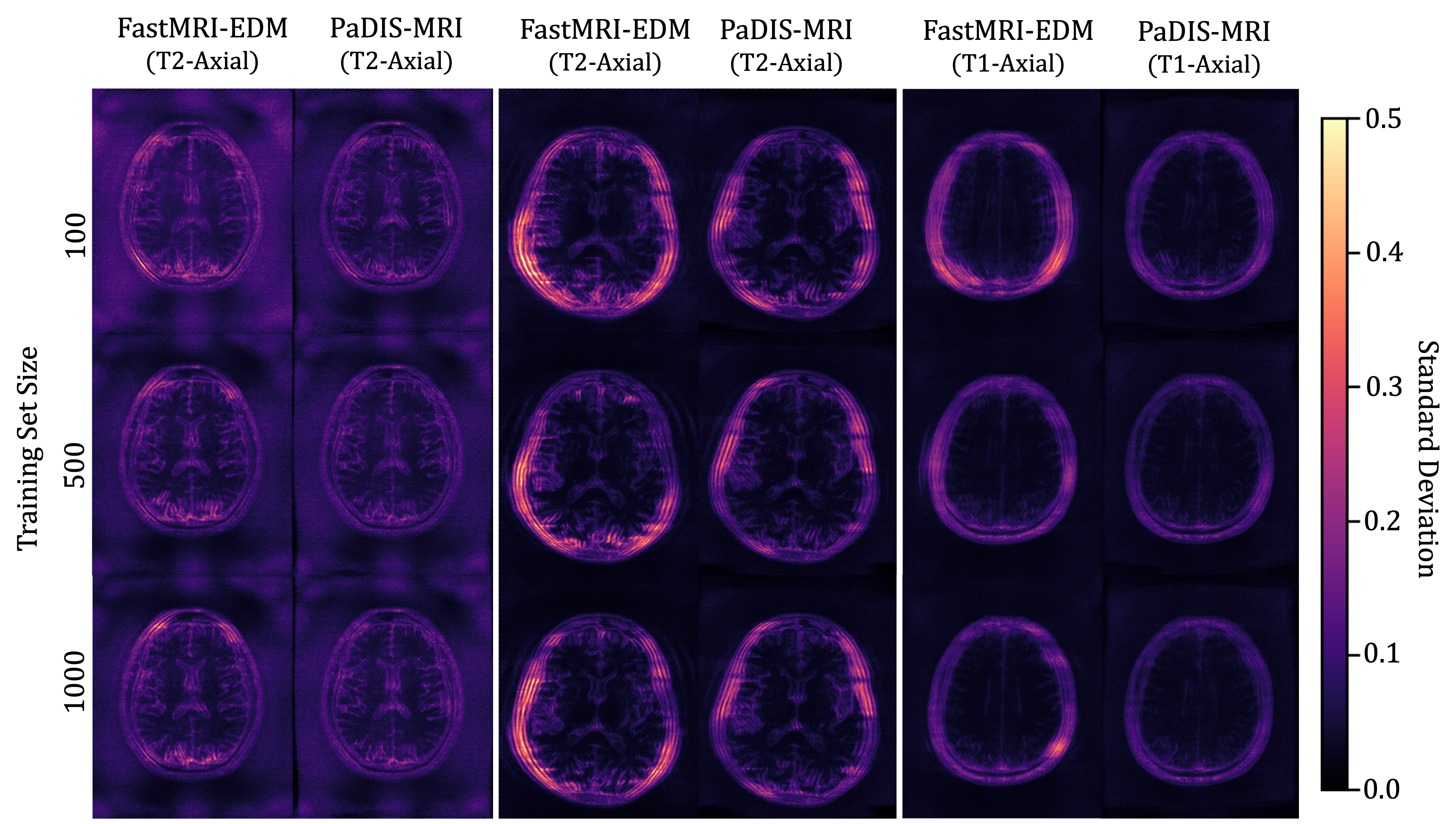}
    \caption{Pixel-wise standard deviations of reconstructions across 10 random k-space sampling masks (\(R=7\)) of the same slice. \textit{Top to Bottom}: models trained on 100, 500, and 1000 slices. PaDIS-MRI consistently exhibits lower pixel-wise variance (fewer bright regions) than FastMRI-EDM across training sizes and contrast types, indicating more stable reconstruction.}
    \vspace{-3mm}
    \label{fig:heatmaps}
\end{figure}

\subsection{Mask-Induced Variability Analysis}
We assess reconstruction consistency by evaluating variability across 10 random k-space undersampling patterns (generated with different random seeds) for the same slice. Since the ground truth image is the same across undersampling masks, lower pixel-wise standard deviation indicates more stable reconstruction. We emphasize this is a consistency metric rather than calibrated posterior uncertainty. Such reliability is of particular interest to clinical settings where higher model consistency paired with superior accuracy can enhance medical decision-making. 

Figure \ref{fig:heatmaps} visualizes these standard deviation maps, with brighter regions indicating higher variability between reconstructions. Across the 100-slice, 500-slice, and 1000-slice training regimes, PaDIS-MRI exhibits noticeably lower pixel-wise standard deviations compared to FastMRI-EDM, with the largest differences occurring at the smaller training dataset sizes. This suggests that decomposing the reconstruction problem into localized patches enables more stable learning of anatomical features, resulting in more reliable and consistent predictions from limited training data.

\begin{table}[ht]
\captionsetup{aboveskip=2pt,belowskip=-6pt}
\setlength{\tabcolsep}{3pt}            
\renewcommand{\arraystretch}{1.08}     
\scriptsize
\centering
\caption{Pixel-wise mean standard deviation (↓) vs.\ training set size $S$.
PICS-L1 is untrained and thus independent of $S$.}
\label{tab:pixelwise_std}
\resizebox{\columnwidth}{!}{%
\begin{tabular}{@{}lcccccc@{}}
\toprule
\multicolumn{1}{c}{Algorithm} & \multicolumn{6}{c}{Training Set Size $S$} \\
\cmidrule(lr){2-7}
& 25 & 100 & 200 & 500 & 1000 & 2330 \\
\midrule
\addlinespace[1pt]
PaDIS\textendash MRI    & \textbf{0.070} & \textbf{0.064} & \textbf{0.063} & \textbf{0.062} & \textbf{0.064} & \textbf{0.062} \\
FastMRI\textendash EDM  & 0.075 & 0.078 & 0.072 & 0.069 & 0.068 & 0.068 \\
PICS\textendash L1      & 0.111     & 0.111     & 0.111     & 0.111     & 0.111     & 0.111 \\
\bottomrule
\end{tabular}%
}
\end{table}

Table \ref{tab:pixelwise_std} quantifies these observations over our evaluation dataset. Due to computational costs of reconstructing each validation image across 10 random seeds, we randomly subsample a smaller evaluation set of 17 images (10 T2-axial, 5 T1-axial, 2 FLAIR) that preserves the contrast-type distribution of the training dataset. We crop each image to its original \(384 \times 320\) dimension to ignore deviations in our zero-padded regions, and report the pixel-wise standard deviation averaged across all test images. Again, PaDIS-MRI consistently achieves lower standard deviation values than FastMRI-EDM on average, with differences being most pronounced at smaller dataset sizes. 
These results further strengthen the case for patch-based diffusion priors in data-scarce settings. A breakdown of pixel-wise standard deviation averages by contrast type can be found in Appendix \ref{appendix:contrast}.

\newcommand{\sigstar}{\textsuperscript{$\star$}} 

\renewcommand{\arraystretch}{1.2}
\setlength{\tabcolsep}{3.5pt}

\begin{table}[ht]
\footnotesize
\centering
\begin{threeparttable}
\caption{Image-level majority preference for PaDIS--MRI (breakdown of the 60 cases). }
\label{tab:reader-pref-inferential}
\begin{tabularx}{\columnwidth}{@{} >{\raggedright\arraybackslash}X r r >{\centering\arraybackslash}p{2.05cm} @{}}
\toprule
Cohort & \multicolumn{1}{c}{\begin{tabular}{@{}c@{}}Picks/\\Total\end{tabular}} & \multicolumn{1}{c}{Prop.} & \multicolumn{1}{c}{95\% CI} \\
\midrule
Majority ($\ge$ 2 of 3 votes) & 55/60 & 91.7\%\sigstar & [81.9, 96.4]\% \\
\addlinespace[2pt]
\multicolumn{4}{@{}l}{\emph{Breakdown by training size $S$ (for data-driven priors)}} \\
\quad 25-slice (n=30)  & 30/30 & 100.0\%\sigstar & [88.7, 100.0]\% \\
\quad 500-slice (n=30) & 25/30 & 83.3\%\sigstar  & [66.4, 92.7]\% \\
\addlinespace[2pt]
\multicolumn{4}{@{}l}{\emph{Breakdown by contrast type}} \\
\quad T1 and FLAIR (n=24)  & 21/24 & 87.5\%\sigstar & [69.0, 95.7]\% \\
\quad T2 (n=36)        & 34/36 & 94.4\%\sigstar & [81.9, 98.5]\% \\
\midrule
Radiologist 1 & 50/60 & 83.3\%\sigstar & [72.0, 90.7]\% \\
Radiologist 2 & 58/60 & 96.7\%\sigstar & [88.6, 99.1]\% \\
Radiologist 3 & 47/60 & 78.3\%\sigstar & [66.4, 86.9]\% \\
\bottomrule
\end{tabularx}
\begin{tablenotes}[flushleft]\footnotesize
\item Proportions are Wilson 95\% CIs. Stars (\sigstar) indicate $p<0.001$ (one-sided exact binomial) significance. 
Image-level rows use $p_0=7/27\approx0.259$; per-radiologist rows use $p_0=1/3\approx0.333$.
\end{tablenotes}
\end{threeparttable}
\end{table}

\begin{table*}[!ht]
\centering
\caption{Reconstruction quality metrics across undersampling ratios with $S=100$ training slices. }
\label{tab:undersample}
\renewcommand{\arraystretch}{1.1}
\setlength{\tabcolsep}{3pt}
\footnotesize

\begin{tabular}{@{}lccc|ccc|ccc@{}}
\toprule
\multirow{2}{*}{$R$} & \multicolumn{3}{c|}{PaDIS-MRI} & \multicolumn{3}{c|}{FastMRI-EDM} & \multicolumn{3}{c}{PICS-L1} \\
 & PSNR $\uparrow$ & SSIM $\uparrow$ & NRMSE $\downarrow$ & PSNR $\uparrow$ & SSIM $\uparrow$ & NRMSE $\downarrow$ & PSNR $\uparrow$ & SSIM $\uparrow$ & NRMSE $\downarrow$ \\
\midrule
2  & 40.34 & 0.898 & 0.062 & \textbf{40.39} & \textbf{0.902} & \textbf{0.061} & 35.45 & 0.748 & 0.106 \\
4  & \textbf{37.24} & 0.862 & \textbf{0.086} & 36.26 & \textbf{0.865} & 0.096 & 29.55 & 0.631 & 0.205 \\
6  & \textbf{35.88} & \textbf{0.864} & \textbf{0.100} & 33.87 & 0.856 & 0.126 & 27.38 & 0.596 & 0.262 \\
8  & \textbf{34.42} & \textbf{0.870} & \textbf{0.118} & 31.61 & 0.846 & 0.164 & 26.25 & 0.582 & 0.298 \\
10 & \textbf{32.82} & \textbf{0.871} & \textbf{0.143} & 30.09 & 0.839 & 0.194 & 25.74 & 0.582 & 0.315 \\
\bottomrule
\end{tabular}

\vspace{2pt}

\scriptsize
\setlength{\tabcolsep}{4pt}
\begin{tabular}{@{}lccc|ccc@{}}
\multicolumn{7}{@{}l}{\emph{Paired differences (mean$\pm$SD); $\Delta$ = PaDIS – EDM (left) and PaDIS – PICS (right).}}\\
\toprule
\multirow{2}{*}{$R$} & \multicolumn{3}{c|}{$\Delta$ vs EDM} & \multicolumn{3}{c}{$\Delta$ vs PICS} \\
 & $\Delta$PSNR & $\Delta$SSIM & $\Delta$NRMSE & $\Delta$PSNR & $\Delta$SSIM & $\Delta$NRMSE \\
\midrule
2  & $-0.05\pm0.56$ & $-0.004\pm0.019$ & $+0.001\pm0.004$ & $+4.89\pm2.11$ & $+0.150\pm0.054$ & $-0.045\pm0.022$ \\
4  & $+0.98\pm0.84$ & $-0.003\pm0.021$ & $-0.010\pm0.009$ & $+7.69\pm2.39$ & $+0.231\pm0.063$ & $-0.118\pm0.038$ \\
6  & $+2.00\pm1.19$ & $+0.007\pm0.025$ & $-0.026\pm0.018$ & $+8.50\pm2.48$ & $+0.267\pm0.066$ & $-0.162\pm0.052$ \\
8  & $+2.81\pm1.09$ & $+0.024\pm0.024$ & $-0.046\pm0.025$ & $+8.17\pm2.43$ & $+0.288\pm0.062$ & $-0.180\pm0.058$ \\
10 & $+2.73\pm1.26$ & $+0.032\pm0.024$ & $-0.051\pm0.028$ & $+7.08\pm2.51$ & $+0.289\pm0.058$ & $-0.173\pm0.062$ \\
\bottomrule
\end{tabular}
\end{table*}

\subsection{Radiologist Verification}
Results of our blinded radiologist preference comparison are summarized in Table \ref{tab:reader-pref-inferential}. Radiologists consistently and significantly preferred PaDIS-MRI across contrast types and training set sizes, with particularly reliable preference for PaDIS-MRI at small training set sizes. Qualitatively, radiologists reported three criteria guiding their choices: (i) \emph{diagnostic interchangeability} with the fully sampled ground truth image, penalizing motion and undersampling artifacts that would make an image unusable; (ii) \emph{clinical accuracy}, no hallucinated structures and no missed fine anatomy (e.g., subtle gyri or small vessels); and (iii) \emph{sharpness}. PaDIS-MRI met these criteria more consistently than either the whole-image FastMRI-EDM diffusion prior or the classical wavelet-sparse reconstruction, likely because PaDIS-MRI's patch-based prior more effectively learns local structural patterns from small datasets. This qualitative feedback from radiologists aligns with PaDIS-MRI's quantitative advantages and better reconstruction consistency, and supports PaDIS-MRI as a promising tool for accelerated MRI reconstruction with limited training data.

To further quantify the strong preference for PaDIS-MRI we compute the diagnostic agreement amongst radiologists. Despite high raw agreement (0.783), Fleiss' $\kappa$ is only 0.105 because the expected-by-chance agreement is correspondingly high (0.758) – illustrating the known $\kappa$-paradox. We therefore interpret $\kappa$ as largely uninformative about consensus; high raw agreement is of greater relevance.

\subsection{Cross-Domain Generalization and Undersampling Robustness} 
\label{cross-contrast}
In clinical settings, models must maintain performance across varying contrast types or modalities that may originate from a distribution different from the one used in training. Additionally, models must demonstrate robustness to different acceleration factors (undersampling rates) during reconstruction. We evaluate PaDIS-MRI in both scenarios.

\renewcommand{\arraystretch}{1.08}

\begin{table}[ht]
\centering
\caption{Cross-Domain generalization: T2-trained models reconstructing T1 and Knee (FS) images at $R=7$.
Arrows indicate direction of improvement. PICS-L1 is untrained.}
\label{tab:cross_contrast_stacked}
\begingroup
\setlength{\tabcolsep}{2.5pt}  
\scriptsize
\begin{tabular}{@{}llcccc@{}}
\toprule
Anatomy & Model & $S$ & PSNR $\uparrow$ & SSIM $\uparrow$ & NRMSE $\downarrow$ \\
\midrule
\multirow{5}{*}{T1}
  & \multirow{2}{*}{PaDIS-MRI}
    & 100 & \textbf{33.90} & \textbf{0.835} & \textbf{0.128} \\
  & & 500 & \textbf{35.02} & \textbf{0.852} & \textbf{0.112} \\
  \cmidrule(lr){2-6}
  & \multirow{2}{*}{FastMRI-EDM}
    & 100 & 31.36 & 0.784 & 0.171 \\
  & & 500 & 31.69 & 0.780 & 0.166 \\
  \cmidrule(lr){2-6}
  & PICS-L1 & -- & 27.73 & 0.587 & 0.261 \\
\midrule
\multirow{5}{*}{Knee (FS)}
  & \multirow{2}{*}{PaDIS-MRI}
    & 100 & \textbf{30.55} & \textbf{0.712} & \textbf{0.182} \\
  & & 500 & \textbf{30.92} & \textbf{0.716} & \textbf{0.175} \\
  \cmidrule(lr){2-6}
  & \multirow{2}{*}{FastMRI-EDM}
    & 100 & 28.58 & 0.647 & 0.227 \\
  & & 500 & 28.11 & 0.640 & 0.239 \\
  \cmidrule(lr){2-6}
  & PICS-L1 & -- & 25.21 & 0.358 & 0.342 \\
\bottomrule
\end{tabular}
\endgroup
\vspace{3pt}
\scriptsize
\centering
\resizebox{\columnwidth}{!}{%
\begin{tabular}{@{}llccc@{}}
\multicolumn{5}{@{}l}{\emph{Paired $\Delta$ (mean$\pm$SD); $\Delta$ = method$_1$–method$_2$. T1: $N{=}32$; Knee: $N{=}30$.}}\\
\toprule
Anatomy & Comparison & $\Delta$PSNR & $\Delta$SSIM & $\Delta$NRMSE \\
\midrule
\multirow{4}{*}{T1}
  & PaDIS–EDM (100)     & $+2.53\pm0.74$ & $+0.051\pm0.031$ & $-0.043\pm0.015$ \\
  & PaDIS–PICS (100)    & $+6.17\pm2.64$ & $+0.248\pm0.090$ & $-0.133\pm0.059$ \\
  & PaDIS–EDM (500)     & $+3.32\pm1.35$ & $+0.072\pm0.040$ & $-0.054\pm0.026$ \\
  & PaDIS–PICS (500)    & $+7.29\pm2.40$ & $+0.265\pm0.090$ & $-0.148\pm0.054$ \\
\midrule
\multirow{4}{*}{Knee}
  & PaDIS–EDM (100)     & $+1.97\pm0.91$ & $+0.065\pm0.027$ & $-0.044\pm0.018$ \\
  & PaDIS–PICS (100)    & $+5.34\pm0.96$ & $+0.354\pm0.044$ & $-0.159\pm0.047$ \\
  & PaDIS–EDM (500)     & $+2.81\pm1.12$ & $+0.076\pm0.033$ & $-0.064\pm0.029$ \\
  & PaDIS–PICS (500)    & $+5.71\pm0.92$ & $+0.358\pm0.043$ & $-0.167\pm0.047$ \\
\bottomrule
\end{tabular}
}
\end{table}

First, we examine cross-contrast generalization by evaluating models trained exclusively on T2-weighted images when reconstructing T1 and FLAIR images at acceleration factor $R=7$, for training set sizes $S \in \{ 100, 500\}$. Table~\ref{tab:cross_contrast_stacked} reveals a significant performance gap, with PaDIS-MRI outperforming both baselines in terms of reconstruction fidelity metrics. PaDIS-MRI also shows similar advantages when these brain-trained models are tested on FS knee MRI, with PaDIS-MRI maintaining strong advantages across all metrics. Notably, PaDIS-MRI exhibits a smaller performance drop between these out-of-distribution experiments and its in-distribution performance (Table~\ref{tab:combined_results} in Appendix \ref{appendix:psnr_ds}) compared to FastMRI-EDM, which appears to require closer alignment between training and evaluation distributions. These cross-domain experiments suggests that PaDIS-MRI's patch-based approach better captures generalizable anatomical features compared to the whole-image model of FastMRI-EDM, and mirrors similar findings by \citet{hu2024patchbaseddiffusionmodelsbeat} in CT reconstruction.

Second, we assess robustness to varying undersampling ratios by testing models trained with $S = 100$ slices across acceleration factors $R \in \{2,4,6,8,10\}$. Table~\ref{tab:undersample} demonstrates that while FastMRI-EDM performs comparably at low acceleration ($R=2$), PaDIS-MRI shows increasingly superior performance as undersampling becomes more severe. 
This widening performance gap suggests that patch-based priors better preserve fine anatomical details when k-space information becomes severely limited, likely because they leverage consistency of local structures rather than relying on global patterns that become increasingly ambiguous under severe undersampling. Appendix \ref{appendix:contrast} contains a breakdown of undersampling rate performance by contrast type.

\section{Discussion}
\label{conclusion}

We extended PaDIS \citep{hu2024learningimagepriorspatchbased} to complex-valued, multi-coil undersampled MRI reconstruction and observed consistent quantitative improvements in fidelity, reconstruction consistency, cross-domain generalization, and robustness compared to a whole-image diffusion prior FastMRI-EDM \citep{aali2024robustmulticoilmrireconstruction}, especially in small-data settings. We then evaluated the accuracy of PaDIS-MRI with a blinded survey in which radiologists consistently found that PaDIS-MRI reconstructions, compared to FastMRI-EDM and a classical sparse-wavelet reconstruction \citep{bart}, are the most similar to fully sampled reference images and therefore more diagnostically useful.

Future work should examine ways to relax the conditional independence assumption inherent in patch-based models, which may limit their ability to capture long-range anatomical dependencies when large training datasets are available. Hybrid approaches that combine patch-based efficiency with mechanisms for modeling global anatomical structure (such as hierarchical patch representations or attention mechanisms across patches) could potentially preserve the data-efficiency advantages of patch-based learning while better leveraging large datasets when available. Balancing the sample-efficiency gains of local patch priors against the need to capture global anatomical context represents an important direction for scaling patch-based diffusion models to diverse clinical applications.

\paragraph{Limitations.}
Although PaDIS-MRI has fewer parameters than FastMRI-EDM (55M vs. 65M), it took roughly twice as long to train. We also observed diminishing marginal improvement with dataset size beyond $\sim$200 slices; this may reflect an earlier ceiling for patch-based priors or limited anatomical diversity from reusing adjacent slices when building larger datasets. These limitations may be addressed by future work to streamline the patch-based training pipeline and adapt it to better leverage large datasets when they are available.

\paragraph{Acknowledgments.} We thank Dr. Jonathan Singh (Department of Radiology, Kaiser Permanente) for serving as a blinded reader in our study. This work was supported in part by the NSF Mathematical Sciences Postdoctoral Research Fellowship under award number 2303178 to SFK. AA is supported by NIH grant R01 HL167974 and ARPA-H contract AY2AX000045.

\clearpage

\bibliography{references}

\clearpage

\appendix
\label{appendix}

\section{Experimental Setup}\label{sec:details}
\paragraph{Datasets.}
\label{appendix:details:datasets}
All experiments use fully sampled k-space data from the NYU Langone Health FastMRI dataset \citep{zbontar2019fastmriopendatasetbenchmarks}, including FLAIR, native T1, pre-contrast T1, post-contrast T1, and axial T2-weighted brain scans. We use the central slice (and additional adjacent slices for larger dataset sizes as discussed below). Following \citet{aali2024robustmulticoilmrireconstruction}, we first apply the inverse Fourier transform to convert the raw multi-coil k-space data into coil images. We estimate the multi-coil noise covariance from a 30 × 30 background patch of coil images and perform pre-whitening of the k-space data using the Berkeley Advanced Reconstruction Toolbox (BART) \citep{bart}. We then normalize the pre-whitened k-space by reconstructing a 24 × 24 centered ACS region using root-sum-of-squares (RSS) and scaling by its 99th percentile. We estimate coil sensitivities from the normalized, pre-whitened central k-space using BART's ESPIRiT. The images are then zero-padded to 384 × 384 pixels (N = 384). For our undersampled validation sets, we apply retrospective Cartesian masks 
at acceleration factors of R \(\in\) {2-10}, fully sample the same 24 × 24 ACS region, randomly select remaining phase-encode lines, and pair each masked k-space with the precomputed coil sensitivity maps. Ground truth images are pre-whitened and normalized, and no extra noise is added beyond the measurement noise intrinsic to FastMRI.

For creating training datasets of various sizes, we sampled slices systematically from randomly selected volumes. For smaller datasets (25, 100, and 200 slices), we extracted only the center slice from each volume to maximize data diversity. For 500 slices, we used two adjacent slices (center and one adjacent) from 250 volumes, while for 1000 slices, we selected four adjacent slices (center plus three adjacent) from 250 volumes. The full dataset of 2330 slices is comprised of five slices from each of the 446 training volumes. All volumes were selected randomly from the training pool without weighting by contrast type. There is no overlap between training and validation volumes from the FastMRI dataset \citep{zbontar2019fastmriopendatasetbenchmarks}. 

In total, the training set contains 446 k-space volumes, consisting of 40 FLAIR, 136 T1-weighted (91 post-contrast, 26 pre-contrast, 19 native), and 270 T2-axial. Our validation set contains 266 volumes (distinct from the training volumes) from which we sample an evaluation test set of center-slices to mirror the training contrast distribution, namely 50 T2-axial, 7 FLAIR, and 25 T1-weighted slices. We also use an evaluation set consisting of 30 knee MRI volumes from the FastMRI dataset (each using the 21st slice) for the cross-modality generalization experiment in Section ~\ref{cross-contrast}.  All metrics are reported over these evaluation test sets.

\paragraph{Models. }
The PaDIS-MRI model consists of approximately 55M parameters and was trained with a padding width of 96 pixels on each side. Patches of sizes \(P \in \{16, 32, 64\}\) were sampled with probabilities \(p \in \{0.2, 0.3, 0.5\}\), respectively. We describe further specific training parameters in Appendix \ref{appendix:traindiff}. We benchmark against the FastMRI-EDM model from \citet{aali2024robustmulticoilmrireconstruction}, a 65M parameter, whole-image diffusion (EDM) model built on the Song-UNet denoiser. Unlike PaDIS-MRI, which trains on zero-padded patches, FastMRI-EDM is evaluated on unpadded, full-resolution images using its original hyperparameters tuned on FastMRI, which align closely with the hyperparameters of PaDIS-MRI (see Appendix \ref{appendix:traindiff}). By preserving each method’s published training regimen, we ensure a fair “out-of-the-box” comparison against a high-performing baseline. Both models were trained until the unconditional generation quality had visually stabilized. We observed that the PaDIS-MRI model took approximately twice as long as the whole-image FastMRI-EDM model to train. We provide detailed training and inference hyperparameters in Section~\ref{appendix:algo}.

\section{Training and Inference Details}
\label{appendix:algo}

Algorithm~\ref{alg:dps2} provides detailed pseudocode for our patch-based VE-DPS algorithm that builds on the PaDIS framework \citep{hu2024learningimagepriorspatchbased} and implements the steps described in Section \ref{sec:methods:dps}. It operates primarily in the padded-image space, only cropping \(x\) to the original field of view when computing the measurement consistency gradient in k-space. The final returned \(x_K\) is also cropped down to the original image dimensions. Note that our implementation of Algorithm \ref{alg:patchdenoising} follows the implementation of Hu et al. \cite{hu2024learningimagepriorspatchbased}.

\newcommand{\RC}[1]{\hfill\texttt{// }#1}

\begin{algorithm*}[!h]
\caption{Patch-based Diffusion Posterior Sampling (\texttt{VE DPS})}
\label{alg:dps2}
\begin{algorithmic}[1]
  \REQUIRE 
    Denoiser $D_\theta: \mathbb{R}^{2\times N'\times N'}\times\mathbb{R}_+\!\to\!\mathbb{R}^{2\times N'\times N'}$,  positional encoding $\text{pos\_enc}$,
    forward operator $\mathcal A$, adjoint $\mathcal A^\dagger$,  
    measurement $y\in\mathbb{C}^{N^2}$,  
    steps $K$, noise bounds $\sigma_{\min},\sigma_{\max}$, exponent $\rho$,  
    data weight $\zeta$, pad $M$, patch size $P$, inner loops $L=10$, $N' = N + 2M$
  \STATE \textbf{Initialize:}
  \STATE \quad $x_0 \gets \mathcal A^\dagger(y)$ \RC{$x_0\in\mathbb{C}^{N\times N}$}
  \STATE \quad $x_0 \gets \text{Pad}(x_0, M)$ \RC{$x_0\in\mathbb{C}^{N'\times N'}$}
  \STATE \textbf{Generate positional encodings:}
  \STATE \quad $x_{\text{grid}}, y_{\text{grid}} \gets \text{linspace}(-1, 1, N')$ \RC{Create normalized coordinate grids}
  \STATE \quad $x_{\text{pos}} \gets$ reshape and repeat $x_{\text{grid}}$ to size $N' \times N'$
  \STATE \quad $y_{\text{pos}} \gets$ reshape and repeat $y_{\text{grid}}$ to size $N' \times N'$ 
  \STATE \quad $\text{pos\_enc} \gets [x_{\text{pos}}, y_{\text{pos}}]$ 
  \STATE \textbf{Precompute timesteps:}
    \[
      t_k \leftarrow \Bigl(\sigma_{\max}^{1/\rho} + \frac{k}{K-1}(\sigma_{\min}^{1/\rho} - \sigma_{\max}^{1/\rho})\Bigr)^\rho,
      \quad
      \alpha_k \leftarrow \frac{1}{2}t_k^2
      \quad (k=0,\dots,K)
    \]
  \FOR{$k=0$ \textbf{to} $K-1$}
    \STATE $x \gets x_k$ \RC{$x\in\mathbb{C}^{N'\times N'}$}
    \FOR{$j=1$ \textbf{to} $L$}
      \STATE Draw $\varepsilon\sim\mathcal N(0,I)$ \RC{Complex Gaussian noise}
      \STATE $\tilde{x} \gets x + t_k\,\varepsilon$ \RC{VE noise injection}
      \STATE $\tilde{x}_{\text{real}} \gets \text{toRealChannels}(\tilde{x})$ \RC{$\tilde{x}_{\text{real}}\in\mathbb{R}^{2\times N'\times N'}$}
      \STATE Randomly sample offset $(a,b) \in [0,M-1]^2$ \RC{For patch extraction}
      \STATE $D_{\text{real}} \gets \text{PatchDenoising}(D_\theta, \tilde{x}_{\text{real}}, t_k, \text{pos\_enc}, P, (a,b))$
      \STATE $D \gets \text{toComplex}(D_{\text{real}})$ \RC{$D\in\mathbb{C}^{N'\times N'}$}
      \STATE $\text{score} \gets (D - x) / t_k^2$ \RC{Score function}
      \STATE $\hat{x} \gets \text{Crop}(D_{\text{real}}, M)$ \RC{$\hat{x}\in\mathbb{R}^{2\times N\times N}$}
      \STATE $r \gets y - \mathcal A(\hat{x})$
      \STATE $\text{SSE} \gets \|r\|^2$
      \STATE $g \gets \nabla_x\,\text{SSE}$ \RC{Compute grad and re-pad back to $g\in\mathbb{C}^{N'\times N'}$}
      \STATE $x \gets x - (\zeta/\sqrt{\text{SSE}})\,g$ \RC{Measurement consistency update}
      \IF{$k < K-1$}
        \STATE $x \gets x + (\alpha_k/2)\,\text{score} + \sqrt{\alpha_k}\,\varepsilon$ 
      \ELSE
        \STATE $x \gets x + (\alpha_k/2)\,\text{score}$ 
      \ENDIF
    \ENDFOR
    \STATE $x_{k+1} \gets x$
  \ENDFOR
  \STATE \textbf{Return} $\text{Crop}(x_K, M)$ \RC{$\in\mathbb{C}^{N\times N}$}
\end{algorithmic}
\end{algorithm*}

\begin{algorithm*}[!h]
\caption{Patch-based Image Denoising \citep{hu2024learningimagepriorspatchbased}}
\label{alg:patchdenoising}
\begin{algorithmic}[1]
  \REQUIRE 
    Neural network $D_\theta$,
    noisy image $x\in\mathbb{R}^{1\times 2 \times N'\times N'}$,
    noise level $t$,
    positional encoding $\text{pos\_enc}$,
    patch size $P$,
    sampling offset $(a,b)$
  \STATE \textbf{Initialize:}
  \STATE \quad $\text{output} \gets \text{zeros\_like}(x)$ \RC{Initialize output tensor}
  \STATE Compute grid points: $\text{grid} \gets \{0, P, 2P, \ldots, (n-1)P\}$ where $n = \lceil N'/P \rceil$
  \STATE Generate patch indices using grid points and offset $(a,b)$:
  \STATE \quad $\text{indices} \gets \{(i+a, i+a+P, j+b, j+b+P) \mid i,j \in \text{grid}\}$
  \STATE $\text{patches} \gets |\text{indices}|$ \RC{Number of patches}
  \STATE Initialize $x_{\text{input}} \in\mathbb{R}^{\text{patches}\times 2 \times P \times P}$
  \STATE Initialize $\text{pos\_input} \in\mathbb{R}^{\text{patches}\times 2 \times P \times P}$
  \FOR{$i=0$ \textbf{to} $\text{patches}-1$}
    \STATE $z \gets \text{indices}[i]$ \RC{$z = (z_1, z_2, z_3, z_4)$ are patch coordinates}
    \STATE $x_{\text{input}}[i] \gets x[0, :, z_1\!:\!z_2, z_3\!:\!z_4]$ \RC{Extract image patch}
    \STATE $\text{pos\_input}[i] \gets \text{pos\_enc}[0, :, z_1\!:\!z_2, z_3\!:\!z_4]$ \RC{Extract positional encodings}
  \ENDFOR
  \STATE $\text{denoised\_patches} \gets D_\theta(x_{\text{input}}, t, \text{pos\_input})$ \RC{Apply denoiser to all patches}
  \FOR{$i=0$ \textbf{to} $\text{patches}-1$}
    \STATE $z \gets \text{indices}[i]$
    \STATE $\text{output}[0, :, z_1\!:\!z_2, z_3\!:\!z_4] \mathrel{+}= \text{denoised\_patches}[i]$ \RC{Accumulate denoised patches}
    \STATE $\text{output}[0, :, z_1\!:\!z_2, z_3\!:\!z_4] \mathrel{-}= x[0, :, z_1\!:\!z_2, z_3\!:\!z_4]$ \RC{Subtract noisy patches}
  \ENDFOR
  \STATE $x_{\text{denoised}} \gets x + \text{output}$ \RC{Add residuals to input}
  \STATE \textbf{Return} $x_{\text{denoised}}$
\end{algorithmic}
\end{algorithm*}

\subsection{Training Hyperparameters}
\label{appendix:traindiff}
The PaDIS-MRI model uses a Song-UNet architecture with standard encoder and decoder components \citep{song2021scorebasedgenerativemodelingstochastic}. It uses 128 base feature channels with multipliers [2,2,2] and a dropout rate of 0.05. Training was performed with a batch size of 4 using the Adam optimizer (learning rate 1e-4, betas [0.9, 0.999]) and was trained using FP32 precision. 
The FastMRI-EDM model was trained on full-size brain MR images using the same Song-UNet architecture type but with a deeper network structure following \citet{aali2024robustmulticoilmrireconstruction}. It uses 128 base channels with multipliers [1,1,2,2,2,2,2], a dropout rate of 0.05, and a batch size of 8. It was trained with a lower learning rate of 5e-5 (with the Adam optimizer and same betas as above) because 1e-4 lead to a higher-variance loss curve. Like PaDIS-MRI, we use FP32 precision. Training was performed on a cluster of NVIDIA RTX A6000s, with each model trained on a single GPU. As discussed in Section \ref{appendix:uncond}, we train models until the unconditional generation visually stabilizes, which is around checkpoint 650 (roughly 2 days) for FastMRI-EDM and checkpoint 1451 (roughly 4.5 days) for PaDIS-MRI. 

\begin{figure}
    \centering
    \includegraphics[width=0.65\linewidth]{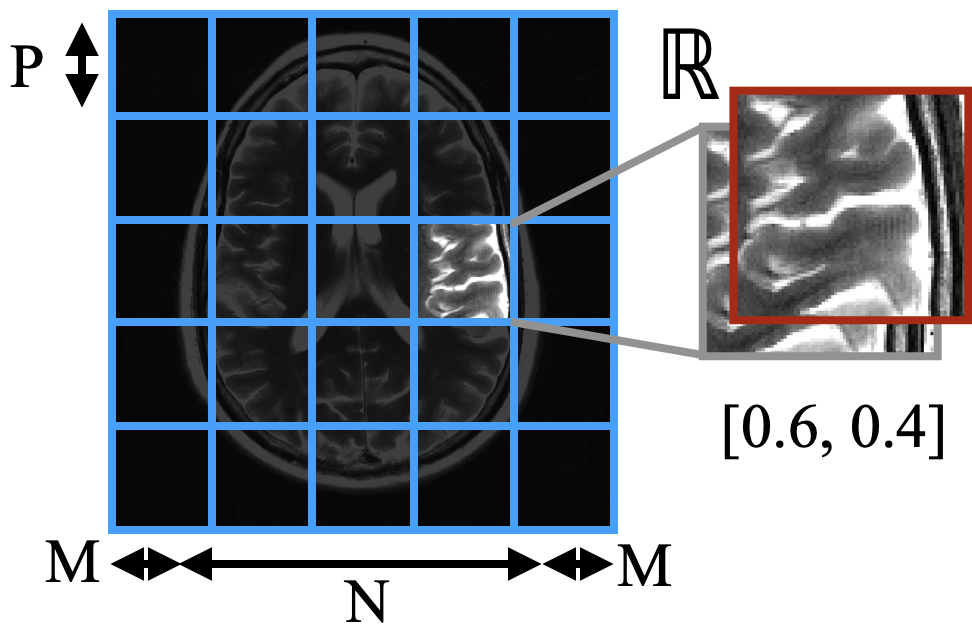}
    \caption{The inputs to our denoising network consist of a two-channel, real-valued representation of a complex-valued patch from the image, along with a positional encoding normalized to [-1, 1].}
    \label{fig:training_visual}
\end{figure}

Figure \ref{fig:training_visual} illustrates how the patch-based training works. The input to our denoising network is a single complex-valued patch that is represented as a two-channel real-valued tensor, with one channel for the real component and one channel for the imaginary component. Following \citet{hu2024learningimagepriorspatchbased}, we pass this tensor to the model along with a 2-dimensional, normalized positional encoding to inform the model as to which part of the full image is represented in the patch. 

\begin{table*}[t]
\centering
\caption{Combined T1/T2/FLAIR reconstruction results for varying training set sizes at undersampling rate $R=7$. Top: Mean $\pm$ SD across $N=82$ slices. Bottom: Paired differences ($\Delta$ = PaDIS-MRI minus comparison method).}
\label{tab:combined_results}
\renewcommand{\arraystretch}{1.1}
\setlength{\tabcolsep}{2pt}
\scriptsize

\begin{tabular}{@{}l|ccc|ccc|ccc|ccc@{}}
\toprule
\multirow{2}{*}{$S$} & \multicolumn{3}{c|}{PaDIS-MRI} & \multicolumn{3}{c|}{FastMRI-EDM} & \multicolumn{3}{c|}{PICS-L1} & \multicolumn{3}{c}{MoDL} \\
 & PSNR$\uparrow$ & SSIM$\uparrow$ & NRMSE$\downarrow$ & PSNR$\uparrow$ & SSIM$\uparrow$ & NRMSE$\downarrow$ & PSNR$\uparrow$ & SSIM$\uparrow$ & NRMSE$\downarrow$ & PSNR$\uparrow$ & SSIM$\uparrow$ & NRMSE$\downarrow$ \\
\midrule
25   & \textbf{32.97{\tiny$\pm$3.4}} & \textbf{.847{\tiny$\pm$.06}} & \textbf{.143{\tiny$\pm$.05}} & 30.51{\tiny$\pm$2.9} & .817{\tiny$\pm$.07} & .185{\tiny$\pm$.05} & 26.76{\tiny$\pm$2.1} & .587{\tiny$\pm$.07} & .281{\tiny$\pm$.04} & 27.86{\tiny$\pm$2.1} & .726{\tiny$\pm$.06} & .247{\tiny$\pm$.05} \\
100  & \textbf{35.05{\tiny$\pm$2.9}} & \textbf{.866{\tiny$\pm$.06}} & \textbf{.110{\tiny$\pm$.03}} & 32.62{\tiny$\pm$3.1} & .853{\tiny$\pm$.07} & .146{\tiny$\pm$.04} & 26.76{\tiny$\pm$2.3} & .587{\tiny$\pm$.08} & .281{\tiny$\pm$.04} & 29.72{\tiny$\pm$1.7} & .752{\tiny$\pm$.06} & .199{\tiny$\pm$.03} \\
200  & \textbf{35.19{\tiny$\pm$3.1}} & \textbf{.866{\tiny$\pm$.07}} & \textbf{.109{\tiny$\pm$.03}} & 33.49{\tiny$\pm$3.0} & .861{\tiny$\pm$.07} & .132{\tiny$\pm$.03} & 26.76{\tiny$\pm$2.4} & .587{\tiny$\pm$.07} & .281{\tiny$\pm$.04} & 30.46{\tiny$\pm$1.7} & .782{\tiny$\pm$.05} & .184{\tiny$\pm$.03} \\
500  & \textbf{35.33{\tiny$\pm$2.9}} & .869{\tiny$\pm$.06} & \textbf{.107{\tiny$\pm$.03}} & 34.34{\tiny$\pm$2.9} & \textbf{.873{\tiny$\pm$.06}} & .119{\tiny$\pm$.03} & 26.76{\tiny$\pm$2.5} & .587{\tiny$\pm$.07} & .281{\tiny$\pm$.03} & 31.46{\tiny$\pm$1.5} & .812{\tiny$\pm$.05} & .165{\tiny$\pm$.04} \\
1000 & \textbf{35.19{\tiny$\pm$2.9}} & .866{\tiny$\pm$.06} & \textbf{.108{\tiny$\pm$.03}} & 34.75{\tiny$\pm$3.0} & \textbf{.876{\tiny$\pm$.06}} & .114{\tiny$\pm$.03} & 26.76{\tiny$\pm$2.5} & .587{\tiny$\pm$.07} & .281{\tiny$\pm$.03} & 32.38{\tiny$\pm$1.3} & .833{\tiny$\pm$.05} & .147{\tiny$\pm$.02} \\
2330 & \textbf{35.31{\tiny$\pm$3.0}} & .865{\tiny$\pm$.07} & \textbf{.107{\tiny$\pm$.03}} & 34.60{\tiny$\pm$2.8} & \textbf{.873{\tiny$\pm$.06}} & .116{\tiny$\pm$.03} & 26.76{\tiny$\pm$2.5} & .587{\tiny$\pm$.07} & .281{\tiny$\pm$.03} & 33.55{\tiny$\pm$1.3} & .859{\tiny$\pm$.05} & .129{\tiny$\pm$.02} \\
\bottomrule
\end{tabular}

\vspace{6pt}

\setlength{\tabcolsep}{2pt}
\begin{tabular}{@{}l|ccc|ccc|ccc@{}}
\multicolumn{10}{@{}l}{\emph{Paired differences (mean$\pm$SD over $N=82$ slices); $\Delta$ = PaDIS-MRI minus comparison method.}}\\
\toprule
\multirow{2}{*}{$S$} 
  & \multicolumn{3}{c|}{$\Delta$ vs EDM} 
  & \multicolumn{3}{c|}{$\Delta$ vs PICS}
  & \multicolumn{3}{c}{$\Delta$ vs MoDL} \\
 & $\Delta$PSNR & $\Delta$SSIM & $\Delta$NRMSE 
 & $\Delta$PSNR & $\Delta$SSIM & $\Delta$NRMSE
 & $\Delta$PSNR & $\Delta$SSIM & $\Delta$NRMSE \\
\midrule
25   & 2.46{\tiny$\pm$1.4} & .031{\tiny$\pm$.03} & $-$.042{\tiny$\pm$.03} & 6.21{\tiny$\pm$3.4} & .261{\tiny$\pm$.07} & $-$.138{\tiny$\pm$.08} & 5.11{\tiny$\pm$2.9} & .121{\tiny$\pm$.06} & $-$.104{\tiny$\pm$.06} \\
100  & 2.43{\tiny$\pm$1.5} & .012{\tiny$\pm$.03} & $-$.036{\tiny$\pm$.03} & 8.28{\tiny$\pm$2.4} & .279{\tiny$\pm$.06} & $-$.171{\tiny$\pm$.05} & 5.33{\tiny$\pm$1.7} & .114{\tiny$\pm$.06} & $-$.089{\tiny$\pm$.03} \\
200  & 1.70{\tiny$\pm$1.3} & .005{\tiny$\pm$.03} & $-$.023{\tiny$\pm$.02} & 8.42{\tiny$\pm$2.5} & .279{\tiny$\pm$.06} & $-$.172{\tiny$\pm$.05} & 4.73{\tiny$\pm$1.7} & .084{\tiny$\pm$.05} & $-$.075{\tiny$\pm$.03} \\
500  & 0.99{\tiny$\pm$0.8} & $-$.004{\tiny$\pm$.02} & $-$.013{\tiny$\pm$.01} & 8.57{\tiny$\pm$2.4} & .282{\tiny$\pm$.06} & $-$.175{\tiny$\pm$.05} & 3.87{\tiny$\pm$1.5} & .057{\tiny$\pm$.05} & $-$.058{\tiny$\pm$.04} \\
1000 & 0.44{\tiny$\pm$1.5} & $-$.010{\tiny$\pm$.02} & $-$.006{\tiny$\pm$.03} & 8.42{\tiny$\pm$2.4} & .279{\tiny$\pm$.06} & $-$.173{\tiny$\pm$.05} & 2.81{\tiny$\pm$1.3} & .033{\tiny$\pm$.05} & $-$.039{\tiny$\pm$.02} \\
2330 & 0.71{\tiny$\pm$0.7} & $-$.008{\tiny$\pm$.02} & $-$.009{\tiny$\pm$.01} & 8.55{\tiny$\pm$2.5} & .279{\tiny$\pm$.07} & $-$.174{\tiny$\pm$.05} & 1.76{\tiny$\pm$1.3} & .006{\tiny$\pm$.05} & $-$.022{\tiny$\pm$.02} \\
\bottomrule
\end{tabular}
\end{table*}

\begin{table*}[!ht]
\centering
\caption{T1/FLAIR MRI reconstruction results for varying training set sizes at undersampling rate $R=7$. Top: Mean $\pm$ SD across $N=32$ slices. Bottom: Paired differences ($\Delta$ = PaDIS-MRI minus comparison method).}
\label{tab:t1flair_results}
\renewcommand{\arraystretch}{1.1}
\setlength{\tabcolsep}{2pt}
\scriptsize

\begin{tabular}{@{}l|ccc|ccc|ccc|ccc@{}}
\toprule
\multirow{2}{*}{$S$} & \multicolumn{3}{c|}{PaDIS-MRI} & \multicolumn{3}{c|}{FastMRI-EDM} & \multicolumn{3}{c|}{PICS-L1} & \multicolumn{3}{c}{MoDL} \\
 & PSNR$\uparrow$ & SSIM$\uparrow$ & NRMSE$\downarrow$ & PSNR$\uparrow$ & SSIM$\uparrow$ & NRMSE$\downarrow$ & PSNR$\uparrow$ & SSIM$\uparrow$ & NRMSE$\downarrow$ & PSNR$\uparrow$ & SSIM$\uparrow$ & NRMSE$\downarrow$ \\
\midrule
25   & \textbf{33.46{\tiny$\pm$2.5}} & \textbf{.840{\tiny$\pm$.06}} & \textbf{.138{\tiny$\pm$.04}} & 31.82{\tiny$\pm$2.6} & .803{\tiny$\pm$.07} & .162{\tiny$\pm$.04} & 27.72{\tiny$\pm$2.2} & .587{\tiny$\pm$.08} & .261{\tiny$\pm$.04} & 29.07{\tiny$\pm$2.1} & .709{\tiny$\pm$.07} & .221{\tiny$\pm$.04} \\
100  & \textbf{36.00{\tiny$\pm$2.4}} & \textbf{.859{\tiny$\pm$.06}} & \textbf{.101{\tiny$\pm$.03}} & 34.01{\tiny$\pm$2.8} & .850{\tiny$\pm$.07} & .128{\tiny$\pm$.03} & 27.72{\tiny$\pm$2.2} & .587{\tiny$\pm$.08} & .261{\tiny$\pm$.04} & 30.77{\tiny$\pm$2.3} & .718{\tiny$\pm$.08} & .182{\tiny$\pm$.03} \\
200  & \textbf{36.31{\tiny$\pm$2.5}} & \textbf{.861{\tiny$\pm$.06}} & \textbf{.097{\tiny$\pm$.03}} & 34.81{\tiny$\pm$2.8} & .855{\tiny$\pm$.07} & .116{\tiny$\pm$.03} & 27.72{\tiny$\pm$2.2} & .587{\tiny$\pm$.08} & .261{\tiny$\pm$.04} & 31.87{\tiny$\pm$2.5} & .760{\tiny$\pm$.08} & .160{\tiny$\pm$.03} \\
500  & \textbf{36.45{\tiny$\pm$2.5}} & .865{\tiny$\pm$.06} & \textbf{.096{\tiny$\pm$.03}} & 35.43{\tiny$\pm$2.7} & \textbf{.869{\tiny$\pm$.06}} & .108{\tiny$\pm$.03} & 27.72{\tiny$\pm$2.2} & .587{\tiny$\pm$.08} & .261{\tiny$\pm$.04} & 32.88{\tiny$\pm$2.5} & .790{\tiny$\pm$.08} & .143{\tiny$\pm$.03} \\
1000 & \textbf{36.31{\tiny$\pm$2.5}} & .863{\tiny$\pm$.06} & \textbf{.097{\tiny$\pm$.03}} & 35.66{\tiny$\pm$2.7} & \textbf{.867{\tiny$\pm$.06}} & .105{\tiny$\pm$.03} & 27.72{\tiny$\pm$2.2} & .587{\tiny$\pm$.08} & .261{\tiny$\pm$.04} & 33.59{\tiny$\pm$2.6} & .812{\tiny$\pm$.08} & .132{\tiny$\pm$.02} \\
2330 & \textbf{36.40{\tiny$\pm$2.5}} & .859{\tiny$\pm$.07} & \textbf{.096{\tiny$\pm$.03}} & 35.69{\tiny$\pm$2.7} & \textbf{.866{\tiny$\pm$.06}} & .104{\tiny$\pm$.03} & 27.72{\tiny$\pm$2.2} & .587{\tiny$\pm$.08} & .261{\tiny$\pm$.04} & 34.86{\tiny$\pm$2.6} & .841{\tiny$\pm$.08} & .114{\tiny$\pm$.02} \\
\bottomrule
\end{tabular}

\vspace{6pt}

\setlength{\tabcolsep}{2pt}
\begin{tabular}{@{}l|ccc|ccc|ccc@{}}
\multicolumn{10}{@{}l}{\emph{Paired differences (mean$\pm$SD over $N=32$ slices); $\Delta$ = PaDIS-MRI minus comparison method.}}\\
\toprule
\multirow{2}{*}{$S$} 
  & \multicolumn{3}{c|}{$\Delta$ vs EDM} 
  & \multicolumn{3}{c|}{$\Delta$ vs PICS}
  & \multicolumn{3}{c}{$\Delta$ vs MoDL} \\
 & $\Delta$PSNR & $\Delta$SSIM & $\Delta$NRMSE 
 & $\Delta$PSNR & $\Delta$SSIM & $\Delta$NRMSE
 & $\Delta$PSNR & $\Delta$SSIM & $\Delta$NRMSE \\
\midrule
25   & 1.64{\tiny$\pm$1.1} & .037{\tiny$\pm$.03} & $-$.025{\tiny$\pm$.02} & 5.74{\tiny$\pm$3.1} & .254{\tiny$\pm$.09} & $-$.123{\tiny$\pm$.07} & 4.39{\tiny$\pm$2.1} & .131{\tiny$\pm$.05} & $-$.084{\tiny$\pm$.04} \\
100  & 1.99{\tiny$\pm$1.5} & .009{\tiny$\pm$.03} & $-$.028{\tiny$\pm$.02} & 8.28{\tiny$\pm$2.6} & .272{\tiny$\pm$.09} & $-$.160{\tiny$\pm$.05} & 5.23{\tiny$\pm$1.7} & .140{\tiny$\pm$.07} & $-$.081{\tiny$\pm$.03} \\
200  & 1.50{\tiny$\pm$1.1} & .006{\tiny$\pm$.04} & $-$.019{\tiny$\pm$.02} & 8.59{\tiny$\pm$2.6} & .274{\tiny$\pm$.09} & $-$.164{\tiny$\pm$.05} & 4.45{\tiny$\pm$1.6} & .101{\tiny$\pm$.07} & $-$.063{\tiny$\pm$.02} \\
500  & 1.02{\tiny$\pm$0.9} & $-$.004{\tiny$\pm$.02} & $-$.012{\tiny$\pm$.01} & 8.77{\tiny$\pm$2.5} & .278{\tiny$\pm$.09} & $-$.165{\tiny$\pm$.05} & 3.57{\tiny$\pm$1.5} & .074{\tiny$\pm$.06} & $-$.047{\tiny$\pm$.02} \\
1000 & 0.65{\tiny$\pm$0.7} & $-$.001{\tiny$\pm$.02} & $-$.008{\tiny$\pm$.01} & 8.58{\tiny$\pm$2.5} & .276{\tiny$\pm$.09} & $-$.164{\tiny$\pm$.05} & 2.72{\tiny$\pm$1.2} & .051{\tiny$\pm$.06} & $-$.035{\tiny$\pm$.02} \\
2330 & 0.71{\tiny$\pm$0.6} & $-$.000{\tiny$\pm$.02} & $-$.008{\tiny$\pm$.01} & 8.67{\tiny$\pm$2.6} & .273{\tiny$\pm$.09} & $-$.165{\tiny$\pm$.06} & 1.54{\tiny$\pm$1.2} & .019{\tiny$\pm$.06} & $-$.018{\tiny$\pm$.01} \\
\bottomrule
\end{tabular}
\end{table*}

\begin{table*}[!ht]
\centering
\caption{T2-weighted MRI reconstruction results for varying training set sizes at undersampling rate $R=7$. Top: Mean $\pm$ SD across $N=50$ slices. Bottom: Paired differences ($\Delta$ = PaDIS-MRI minus comparison method).}
\label{tab:t2_results}
\renewcommand{\arraystretch}{1.1}
\setlength{\tabcolsep}{2pt}
\scriptsize

\begin{tabular}{@{}l|ccc|ccc|ccc|ccc@{}}
\toprule
\multirow{2}{*}{$S$} & \multicolumn{3}{c|}{PaDIS-MRI} & \multicolumn{3}{c|}{FastMRI-EDM} & \multicolumn{3}{c|}{PICS-L1} & \multicolumn{3}{c}{MoDL} \\
 & PSNR$\uparrow$ & SSIM$\uparrow$ & NRMSE$\downarrow$ & PSNR$\uparrow$ & SSIM$\uparrow$ & NRMSE$\downarrow$ & PSNR$\uparrow$ & SSIM$\uparrow$ & NRMSE$\downarrow$ & PSNR$\uparrow$ & SSIM$\uparrow$ & NRMSE$\downarrow$ \\
\midrule
25   & \textbf{32.66{\tiny$\pm$3.2}} & \textbf{.852{\tiny$\pm$.06}} & \textbf{.146{\tiny$\pm$.05}} & 29.67{\tiny$\pm$2.8} & .826{\tiny$\pm$.06} & .200{\tiny$\pm$.06} & 26.15{\tiny$\pm$2.0} & .587{\tiny$\pm$.07} & .294{\tiny$\pm$.04} & 27.09{\tiny$\pm$1.7} & .738{\tiny$\pm$.06} & .264{\tiny$\pm$.04} \\
100  & \textbf{34.44{\tiny$\pm$2.8}} & \textbf{.869{\tiny$\pm$.05}} & \textbf{.116{\tiny$\pm$.04}} & 31.73{\tiny$\pm$2.8} & .856{\tiny$\pm$.06} & .157{\tiny$\pm$.05} & 26.15{\tiny$\pm$2.0} & .587{\tiny$\pm$.07} & .294{\tiny$\pm$.04} & 29.05{\tiny$\pm$1.9} & .772{\tiny$\pm$.07} & .211{\tiny$\pm$.03} \\
200  & \textbf{34.47{\tiny$\pm$2.9}} & \textbf{.869{\tiny$\pm$.05}} & \textbf{.117{\tiny$\pm$.04}} & 32.64{\tiny$\pm$2.8} & .865{\tiny$\pm$.06} & .141{\tiny$\pm$.04} & 26.15{\tiny$\pm$2.0} & .587{\tiny$\pm$.07} & .294{\tiny$\pm$.04} & 29.56{\tiny$\pm$1.9} & .796{\tiny$\pm$.06} & .199{\tiny$\pm$.03} \\
500  & \textbf{34.62{\tiny$\pm$2.7}} & .872{\tiny$\pm$.05} & \textbf{.113{\tiny$\pm$.03}} & 33.65{\tiny$\pm$2.7} & \textbf{.875{\tiny$\pm$.05}} & .126{\tiny$\pm$.03} & 26.15{\tiny$\pm$2.0} & .587{\tiny$\pm$.07} & .294{\tiny$\pm$.04} & 30.56{\tiny$\pm$2.0} & .826{\tiny$\pm$.06} & .178{\tiny$\pm$.03} \\
1000 & \textbf{34.47{\tiny$\pm$2.7}} & .869{\tiny$\pm$.05} & \textbf{.116{\tiny$\pm$.04}} & 34.17{\tiny$\pm$2.8} & \textbf{.881{\tiny$\pm$.05}} & .120{\tiny$\pm$.04} & 26.15{\tiny$\pm$2.0} & .587{\tiny$\pm$.07} & .294{\tiny$\pm$.04} & 31.60{\tiny$\pm$2.0} & .846{\tiny$\pm$.06} & .157{\tiny$\pm$.03} \\
2330 & \textbf{34.62{\tiny$\pm$2.8}} & .869{\tiny$\pm$.05} & \textbf{.114{\tiny$\pm$.04}} & 33.90{\tiny$\pm$2.7} & \textbf{.879{\tiny$\pm$.05}} & .123{\tiny$\pm$.04} & 26.15{\tiny$\pm$2.0} & .587{\tiny$\pm$.07} & .294{\tiny$\pm$.04} & 32.72{\tiny$\pm$1.9} & .872{\tiny$\pm$.06} & .138{\tiny$\pm$.02} \\
\bottomrule
\end{tabular}

\vspace{6pt}

\setlength{\tabcolsep}{2pt}
\begin{tabular}{@{}l|ccc|ccc|ccc@{}}
\multicolumn{10}{@{}l}{\emph{Paired differences (mean$\pm$SD over $N=50$ slices); $\Delta$ = PaDIS-MRI minus comparison method.}}\\
\toprule
\multirow{2}{*}{$S$} 
  & \multicolumn{3}{c|}{$\Delta$ vs EDM} 
  & \multicolumn{3}{c|}{$\Delta$ vs PICS}
  & \multicolumn{3}{c}{$\Delta$ vs MoDL} \\
 & $\Delta$PSNR & $\Delta$SSIM & $\Delta$NRMSE 
 & $\Delta$PSNR & $\Delta$SSIM & $\Delta$NRMSE
 & $\Delta$PSNR & $\Delta$SSIM & $\Delta$NRMSE \\
\midrule
25   & 2.99{\tiny$\pm$1.4} & .026{\tiny$\pm$.02} & $-$.054{\tiny$\pm$.03} & 6.51{\tiny$\pm$3.5} & .265{\tiny$\pm$.05} & $-$.148{\tiny$\pm$.08} & 5.57{\tiny$\pm$3.2} & .115{\tiny$\pm$.06} & $-$.117{\tiny$\pm$.07} \\
100  & 2.71{\tiny$\pm$1.4} & .014{\tiny$\pm$.02} & $-$.041{\tiny$\pm$.03} & 8.29{\tiny$\pm$2.3} & .283{\tiny$\pm$.04} & $-$.178{\tiny$\pm$.05} & 5.39{\tiny$\pm$1.7} & .098{\tiny$\pm$.04} & $-$.095{\tiny$\pm$.03} \\
200  & 1.83{\tiny$\pm$1.4} & .004{\tiny$\pm$.02} & $-$.025{\tiny$\pm$.02} & 8.32{\tiny$\pm$2.4} & .282{\tiny$\pm$.04} & $-$.178{\tiny$\pm$.05} & 4.91{\tiny$\pm$1.8} & .073{\tiny$\pm$.04} & $-$.082{\tiny$\pm$.03} \\
500  & 0.98{\tiny$\pm$0.8} & $-$.004{\tiny$\pm$.01} & $-$.013{\tiny$\pm$.01} & 8.47{\tiny$\pm$2.3} & .285{\tiny$\pm$.04} & $-$.181{\tiny$\pm$.05} & 4.06{\tiny$\pm$1.5} & .046{\tiny$\pm$.03} & $-$.064{\tiny$\pm$.02} \\
1000 & 0.31{\tiny$\pm$1.9} & $-$.013{\tiny$\pm$.02} & $-$.005{\tiny$\pm$.03} & 8.32{\tiny$\pm$2.3} & .281{\tiny$\pm$.04} & $-$.179{\tiny$\pm$.05} & 2.87{\tiny$\pm$1.3} & .022{\tiny$\pm$.03} & $-$.042{\tiny$\pm$.02} \\
2330 & 0.71{\tiny$\pm$0.8} & $-$.009{\tiny$\pm$.01} & $-$.009{\tiny$\pm$.01} & 8.47{\tiny$\pm$2.3} & .282{\tiny$\pm$.04} & $-$.181{\tiny$\pm$.05} & 1.89{\tiny$\pm$1.3} & $-$.002{\tiny$\pm$.03} & $-$.025{\tiny$\pm$.02} \\
\bottomrule
\end{tabular}
\end{table*}

\subsection{Inference Hyperparameters} 
Our VE-DPS algorithms for both FastMRI-EDM and PaDIS-MRI use a shared noise schedule over $K = 104$ steps, with bounds $\sigma_{min} = 0.003$ to $\sigma_{max} = 10.0$ and no churn ($S_\text{churn} = 0$). In the patch-based implementation, we explicitly loop through the 104 noise levels with 10 inner iterations per level, while in the whole-image implementation, we equivalently construct a 1040-step trajectory where each noise level is repeated 10 times. For the measurement update step, we use a data fidelity weight $\zeta = 3.0$, pad by $M=64$, and use patch size $P=64$. Inference was also run on our NVIDIA RTX A6000 GPUs. Examples of each model's unconditional generations can be found in Appendix \ref{appendix:uncond}.

\subsection{Unconditional Generation Quality}
\label{appendix:uncond}
\begin{figure}
    \centering
    \includegraphics[width=1.0\linewidth]{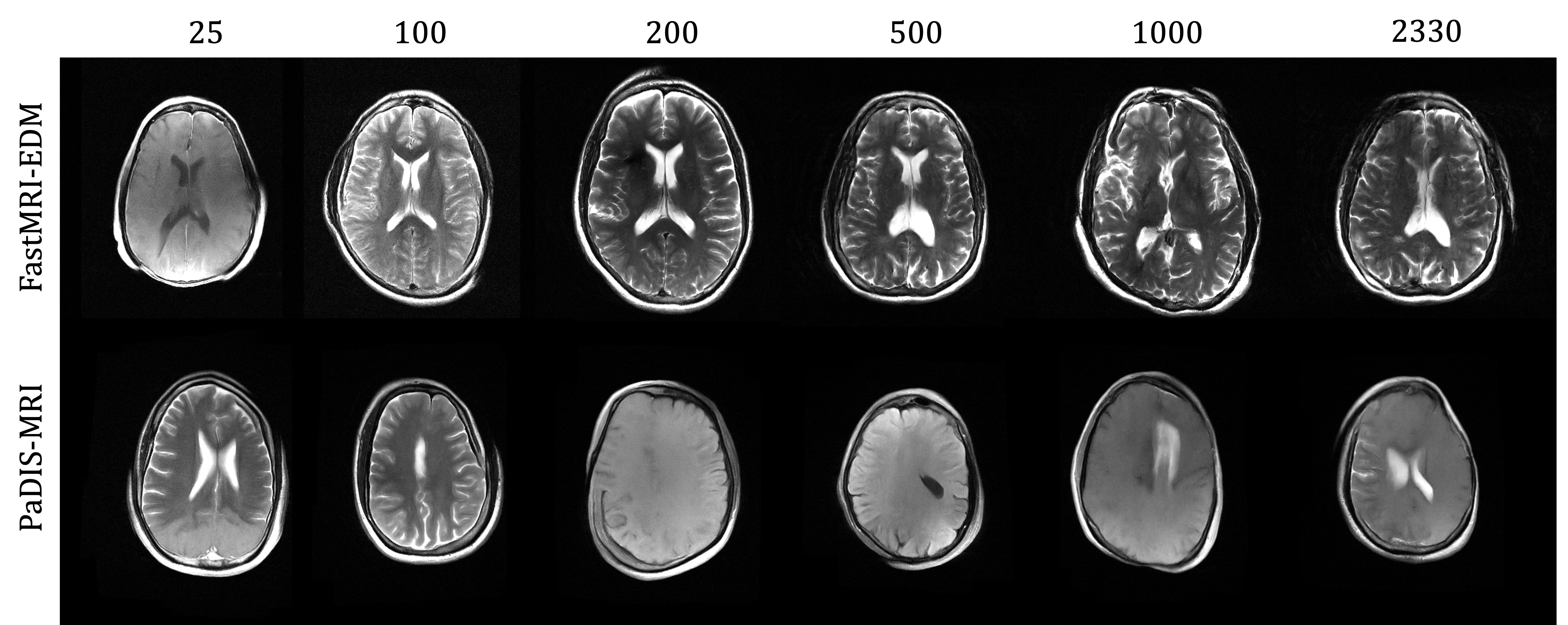}
    \caption{Unconditional image generation quality comparison between FastMRI-EDM (top row) and PaDIS-MRI (bottom row) across different training dataset sizes (25 to 2330 slices, indicated above each column). PaDIS-MRI produces less visually intricate unconditional generations, yet provides a more reliable prior for conditional generation in undersampled MRI reconstruction.
    }
    \label{fig:uncond_generation_comparison}
\end{figure}

Figure~\ref{fig:uncond_generation_comparison} shows sampled unconditional generations for both PaDIS-MRI and FastMRI-EDM, at model convergence. As discussed previously, we trained models by qualitatively evaluating the change in unconditional generation quality over time. The FastMRI-EDM model is able to capture more detailed anatomy in its unconditional generations, whereas PaDIS-MRI tends to generate more uniformly gray ovals with limited detail. We hypothesize that the relatively poorer-looking unconditional generations of PaDIS-MRI may actually make it more robust during inference, or conditional generation. Rather than capturing intricate global features, the relatively simple structure may provide a form of implicit regularization that favors locally consistent features over potentially spurious or overfitted global patterns. 

The training dynamics also differ substantially between the two approaches, with PaDIS-MRI requiring approximately 1450 checkpoints to stabilize compared to 650 for FastMRI-EDM. We expect further development of patch-based models may reduce this overhead in training time. Nonetheless, this gap in training time does not explain the PaDIS-MRI model's superior performance at small dataset sizes. Several experiments were done with the FastMRI-EDM at checkpoints of 1200 and 1450; the performance of FastMRI-EDM was worse at these checkpoints than its performance at checkpoint 650. Thus, we do not believe that FastMRI-EDM was undertrained relative to PaDIS-MRI.

\subsection{Supervised Baseline (MoDL) Implementation}
\label{appendix:modl}
We implement MoDL following \citet{modl}. The forward operator uses SENSE with precomputed coil sensitivity maps, and each unrolled iteration alternates (i) a U\hbox{-}Net prior module and (ii) a conjugate-gradient (CG) data-consistency (DC) update. We unroll six DC steps. The DC weight $\lambda$ is a learned scalar initialized to $0.1$. Models are trained on magnitude images with an NRMSE loss for 10 epochs using batch size~1; other optimizer and preprocessing choices match Appendix~\ref{sec:details}. At test time, MoDL is evaluated with the same k\hbox{-}space masks and metrics as our diffusion baselines. This implementation follows that used in ~\citet{aali2024robustmulticoilmrireconstruction}.

\section{Additional Experimental Results}
\subsection{Dataset Size}
\label{appendix:psnr_ds}

We present the tabular form of Figure \ref{fig:psnr_ds} here in Table \ref{tab:combined_results}. Across reconstruction quality metrics, PaDIS-MRI outperforms FastMRI-EDM, MoDL, and PICS-L1, especially at small dataset sizes. Moreover, the paired difference statistics confirm that the quantitative improvements of PaDIS-MRI are consistent across images in the dataset. 

Tables~\ref{tab:t1flair_results} and~\ref{tab:t2_results} provide a contrast-specific breakdown of reconstruction performance across training dataset sizes, complementing the combined results presented in Figure \ref{fig:psnr_ds} and Table \ref{tab:combined_results}. 
For T1 and FLAIR contrasts (Table~\ref{tab:t1flair_results}), PaDIS-MRI consistently outperforms FastMRI-EDM across all dataset sizes, with particularly pronounced advantages at smaller dataset sizes.
For T2-weighted images (Table~\ref{tab:t2_results}), we observe a similar trend with PaDIS-MRI showing substantial improvements at smaller dataset sizes.
Interestingly, while PaDIS-MRI maintains higher PSNR and lower NRMSE at larger dataset sizes, FastMRI-EDM achieves marginally better SSIM scores for T2 reconstructions at $S \geq 500$. This suggests that the whole-image approach may better preserve certain structural features in T2 images when sufficient training data is available, despite its overall lower fidelity in terms of pixel-wise error.

\subsection{Mask-Induced Variability}
\label{appendix:contrast}

Table~\ref{tab:pixelwise_std_contrast} presents the pixel-wise standard deviation results broken down by contrast type. 
For T2-weighted images, PaDIS-MRI consistently demonstrates lower pixel-wise standard deviation than FastMRI-EDM across all dataset sizes, with differences being most pronounced at smaller dataset sizes.
For T1 and FLAIR contrasts, the advantage of PaDIS-MRI becomes evident at $S \geq 100$, with substantial reductions in standard deviation compared to FastMRI-EDM. Interestingly, at the smallest dataset size ($S=25$), FastMRI-EDM shows slightly lower uncertainty for T1/FLAIR contrasts despite its poorer reconstruction quality, perhaps because at this level its errors are dominated by bias rather than variance. 

\begin{table*}[!h]
\centering
\caption{Pixel-wise standard deviation (\(\downarrow\)) by contrast type across dataset sizes at undersampling rate $R=7$. }
\label{tab:pixelwise_std_contrast}
\renewcommand{\arraystretch}{1.1}
\setlength{\tabcolsep}{3pt}
\footnotesize
\begin{tabular}{c|cc|cc|cc}
\toprule
\multirow{2}{*}{Dataset Size ($S$)} & \multicolumn{2}{c|}{PaDIS-MRI} & \multicolumn{2}{c|}{FastMRI-EDM} & \multicolumn{2}{c}{PICS-L1} \\
 & T2 & T1 & T2 & T1 & T2 & T1 \\
\midrule
25   & \textbf{0.069} & 0.073 & 0.078 & \textbf{0.070} & 0.114 & 0.107 \\
100  & \textbf{0.070} & \textbf{0.056} & 0.086 & 0.065 & 0.114 & 0.107 \\
200  & \textbf{0.070} & \textbf{0.053} & 0.080 & 0.061 & 0.114 & 0.107 \\
500  & \textbf{0.069} & \textbf{0.053} & 0.077 & 0.058 & 0.114 & 0.107 \\
1000 & \textbf{0.072} & \textbf{0.054} & 0.076 & 0.056 & 0.114 & 0.107 \\
2330 & \textbf{0.069} & \textbf{0.052} & 0.076 & 0.056 & 0.114 & 0.107 \\
\bottomrule
\end{tabular}
\end{table*}

\subsection{Robustness}
Tables~\ref{tab:t1_undersample} and~\ref{tab:t2_undersample} provide contrast-specific performance across different undersampling ratios for models trained with $S=100$ slices. For T1 and FLAIR contrasts (Table~\ref{tab:t1_undersample}), FastMRI-EDM shows marginally better performance at the lowest acceleration factor ($R=2$), but PaDIS-MRI quickly establishes dominance as the acceleration factor increases. 

A similar pattern emerges for T2-weighted images (Table~\ref{tab:t2_undersample}), where FastMRI-EDM performs slightly better at low acceleration factors ($R=2$) and maintains a higher SSIM at $R=4$ despite lower PSNR. However, as undersampling becomes more severe ($R \geq 6$), PaDIS-MRI clearly outperforms FastMRI-EDM across all metrics. 

These contrast-specific results reinforce our finding that patch-based priors offer improved robustness to severe undersampling, with the advantage becoming increasingly pronounced at higher acceleration factors as the reconstruction relies more heavily on the diffusion prior. This pattern holds true across different anatomical contrasts, though the magnitude of improvement varies between T1 and T2 weighted images.

\begin{table*}[ht]
\centering
\caption{T1/FLAIR reconstruction quality metrics across undersampling ratios with $S=100$ training slices.}
\label{tab:t1_undersample}
\renewcommand{\arraystretch}{1.1}
\setlength{\tabcolsep}{3pt}
\footnotesize

\begin{tabular}{@{}lccc|ccc|ccc@{}}
\toprule
\multirow{2}{*}{$R$} & \multicolumn{3}{c|}{PaDIS-MRI} & \multicolumn{3}{c|}{FastMRI-EDM} & \multicolumn{3}{c}{PICS\textendash L1} \\
 & PSNR $\uparrow$ & SSIM $\uparrow$ & NRMSE $\downarrow$ & PSNR $\uparrow$ & SSIM $\uparrow$ & NRMSE $\downarrow$ & PSNR $\uparrow$ & SSIM $\uparrow$ & NRMSE $\downarrow$ \\
\midrule
2  & 40.92 & \textbf{0.888} & 0.059 & \textbf{40.94} & 0.887 & \textbf{0.058} & 35.85 & 0.746 & 0.105 \\
4  & \textbf{37.88} & \textbf{0.852} & \textbf{0.082} & 37.19 & 0.851 & 0.089 & 30.06 & 0.618 & 0.201 \\
6  & \textbf{36.57} & \textbf{0.854} & \textbf{0.095} & 34.84 & 0.845 & 0.116 & 28.28 & 0.593 & 0.243 \\
8  & \textbf{35.42} & \textbf{0.866} & \textbf{0.108} & 33.10 & 0.847 & 0.141 & 27.29 & 0.587 & 0.273 \\
10 & \textbf{33.95} & \textbf{0.871} & \textbf{0.129} & 31.54 & 0.845 & 0.169 & 26.73 & 0.593 & 0.291 \\
\bottomrule
\end{tabular}

\vspace{2pt}

\scriptsize
\setlength{\tabcolsep}{4pt}
\begin{tabular}{@{}lccc|ccc@{}}
\multicolumn{7}{@{}l}{\emph{Paired differences (mean$\pm$SD; $N{=}32$); $\Delta$ = PaDIS $-$ EDM (left) and PaDIS $-$ PICS (right).}}\\
\toprule
\multirow{2}{*}{$R$} & \multicolumn{3}{c|}{$\Delta$ vs EDM} & \multicolumn{3}{c}{$\Delta$ vs PICS} \\
 & $\Delta$PSNR & $\Delta$SSIM & $\Delta$NRMSE & $\Delta$PSNR & $\Delta$SSIM & $\Delta$NRMSE \\
\midrule
2  & $-0.02\pm0.71$ & $+0.001\pm0.025$ & $+0.001\pm0.005$ & $+5.07\pm2.54$ & $+0.142\pm0.070$ & $-0.046\pm0.028$ \\
4  & $+0.70\pm0.90$ & $+0.001\pm0.028$ & $-0.006\pm0.008$ & $+7.82\pm2.41$ & $+0.234\pm0.089$ & $-0.118\pm0.041$ \\
6  & $+1.72\pm1.39$ & $+0.009\pm0.033$ & $-0.022\pm0.021$ & $+8.28\pm2.28$ & $+0.261\pm0.091$ & $-0.148\pm0.043$ \\
8  & $+2.32\pm1.25$ & $+0.019\pm0.032$ & $-0.033\pm0.019$ & $+8.13\pm2.40$ & $+0.279\pm0.084$ & $-0.166\pm0.056$ \\
10 & $+2.41\pm1.35$ & $+0.026\pm0.029$ & $-0.040\pm0.026$ & $+7.21\pm2.74$ & $+0.278\pm0.079$ & $-0.161\pm0.062$ \\
\bottomrule
\end{tabular}
\end{table*}

\begin{table*}[ht]
\centering
\caption{T2 reconstruction quality metrics across undersampling ratios with $S=100$ training slices.}
\label{tab:t2_undersample}
\renewcommand{\arraystretch}{1.1}
\setlength{\tabcolsep}{3pt}
\footnotesize

\begin{tabular}{@{}lccc|ccc|ccc@{}}
\toprule
\multirow{2}{*}{$R$} & \multicolumn{3}{c|}{PaDIS-MRI} & \multicolumn{3}{c|}{FastMRI-EDM} & \multicolumn{3}{c}{PICS-L1} \\
 & PSNR $\uparrow$ & SSIM $\uparrow$ & NRMSE $\downarrow$ & PSNR $\uparrow$ & SSIM $\uparrow$ & NRMSE $\downarrow$ & PSNR $\uparrow$ & SSIM $\uparrow$ & NRMSE $\downarrow$ \\
\midrule
2  & 39.97 & 0.905 & 0.063 & \textbf{40.03} & \textbf{0.911} & \textbf{0.062} & 35.19 & 0.750 & 0.107 \\
4  & \textbf{36.83} & 0.869 & \textbf{0.089} & 35.67 & \textbf{0.874} & 0.101 & 29.23 & 0.639 & 0.208 \\
6  & \textbf{35.44} & \textbf{0.870} & \textbf{0.103} & 33.25 & 0.863 & 0.131 & 26.80 & 0.598 & 0.274 \\
8  & \textbf{33.78} & \textbf{0.873} & \textbf{0.124} & 30.65 & 0.846 & 0.179 & 25.58 & 0.578 & 0.314 \\
10 & \textbf{32.10} & \textbf{0.871} & \textbf{0.151} & 29.17 & 0.835 & 0.210 & 25.11 & 0.575 & 0.331 \\
\bottomrule
\end{tabular}

\vspace{2pt}

\scriptsize
\setlength{\tabcolsep}{4pt}
\begin{tabular}{@{}lccc|ccc@{}}
\multicolumn{7}{@{}l}{\emph{Paired differences (mean$\pm$SD; N=50); $\Delta$ = PaDIS – EDM (left) and PaDIS – PICS (right).}}\\
\toprule
\multirow{2}{*}{$R$} & \multicolumn{3}{c|}{$\Delta$ vs EDM} & \multicolumn{3}{c}{$\Delta$ vs PICS} \\
 & $\Delta$PSNR & $\Delta$SSIM & $\Delta$NRMSE & $\Delta$PSNR & $\Delta$SSIM & $\Delta$NRMSE \\
\midrule
2  & $-0.06\pm0.44$ & $-0.007\pm0.012$ & $+0.001\pm0.004$ & $+4.77\pm1.77$ & $+0.155\pm0.041$ & $-0.044\pm0.018$ \\
4  & $+1.16\pm0.75$ & $-0.006\pm0.015$ & $-0.012\pm0.008$ & $+7.60\pm2.38$ & $+0.230\pm0.038$ & $-0.119\pm0.035$ \\
6  & $+2.19\pm1.00$ & $+0.007\pm0.017$ & $-0.029\pm0.015$ & $+8.64\pm2.58$ & $+0.271\pm0.043$ & $-0.172\pm0.056$ \\
8  & $+3.12\pm0.85$ & $+0.027\pm0.016$ & $-0.055\pm0.025$ & $+8.20\pm2.44$ & $+0.294\pm0.041$ & $-0.190\pm0.057$ \\
10 & $+2.93\pm1.16$ & $+0.036\pm0.019$ & $-0.059\pm0.026$ & $+6.99\pm2.35$ & $+0.297\pm0.037$ & $-0.180\pm0.062$ \\
\bottomrule
\end{tabular}
\end{table*}

\newcommand\mrizoom[3]{%
  \begin{tikzpicture}[
    zoomboxarray,
    zoomboxes below,
    connect zoomboxes,
    zoombox paths/.append style={thick}
  ]
    \node[image node]{\includegraphics[width=0.19\textwidth]{#1}};
    \zoombox[magnification=4,color code=MRIcolA]{#2}
    \zoombox[magnification=4,color code=MRIcolB]{#3}
  \end{tikzpicture}%
}
\begin{figure*}[p]
  \centering
  \setlength{\tabcolsep}{0pt}
  \def\arraystretch{0.5}%
  \begin{tabular}{@{}c@{}c@{}c@{}c@{}c@{}}
    \multicolumn{1}{c}{T2-Axial} &
    \multicolumn{1}{c}{T2-Axial} &
    \multicolumn{1}{c}{T2-Axial} &
    \multicolumn{1}{c}{FLAIR} &
    \multicolumn{1}{c}{T1} \\[5pt]
    
    \mrizoom{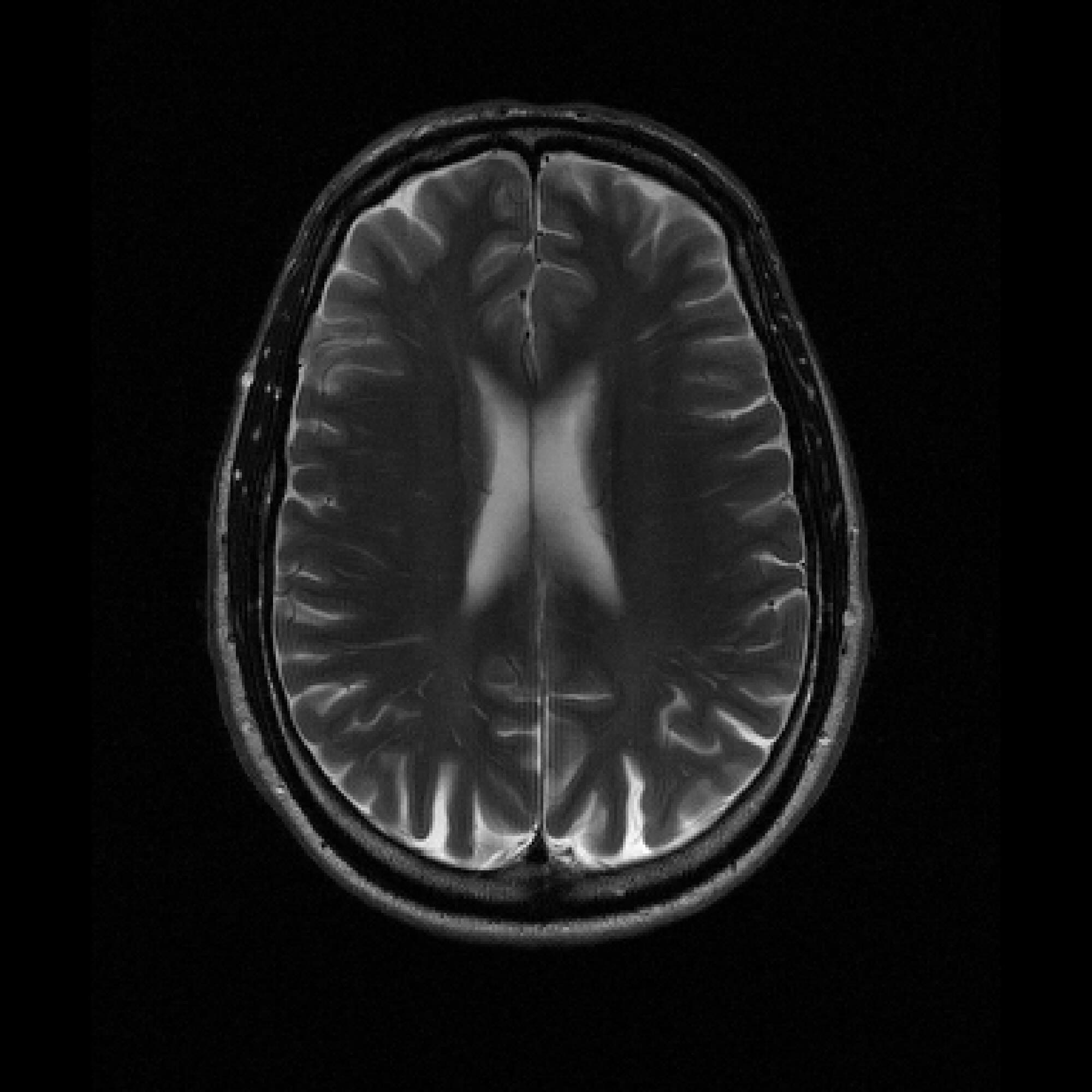}    {0.28,0.44} {0.47,0.25} &
    \mrizoom{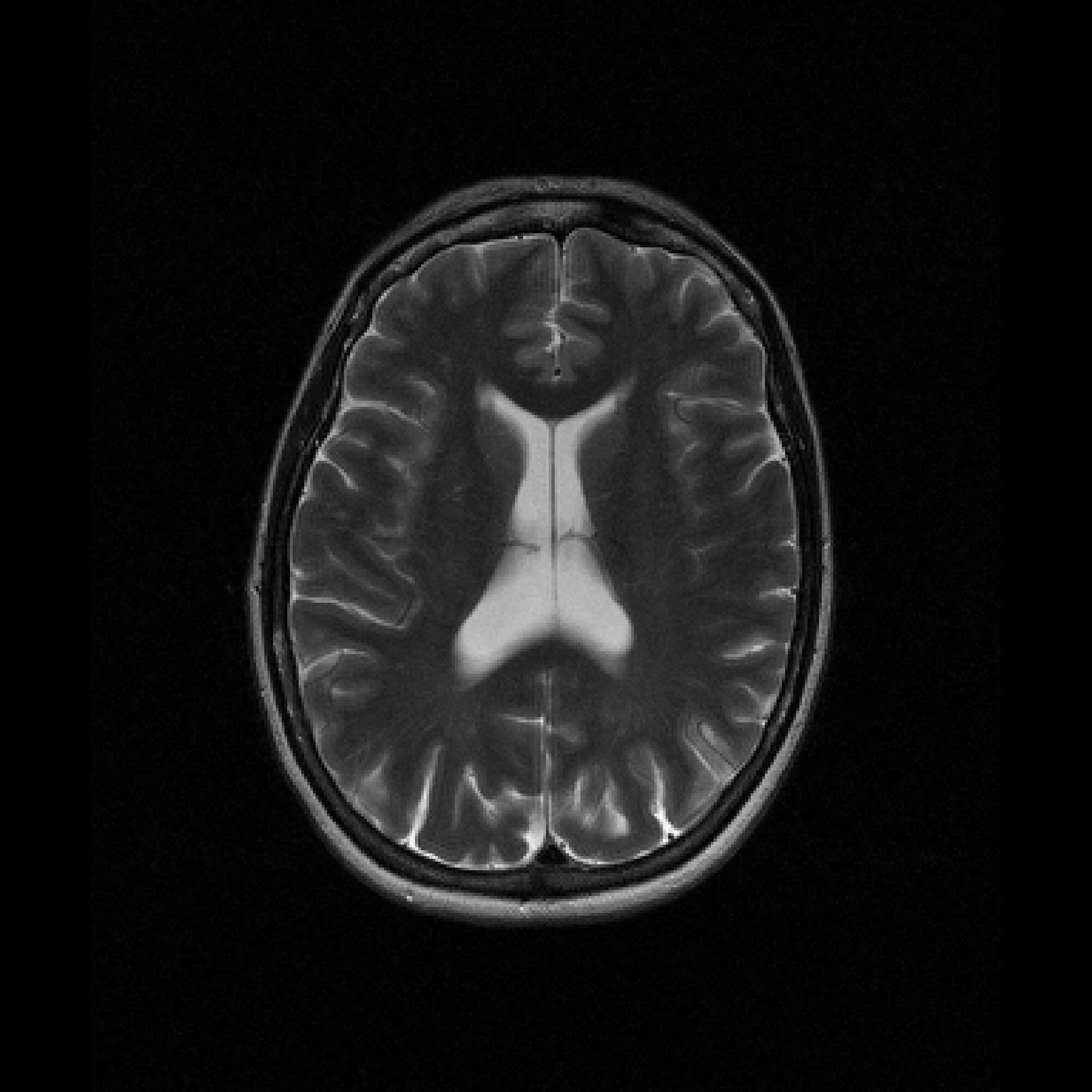}      {0.35,0.45} {0.5,0.25} &
    \mrizoom{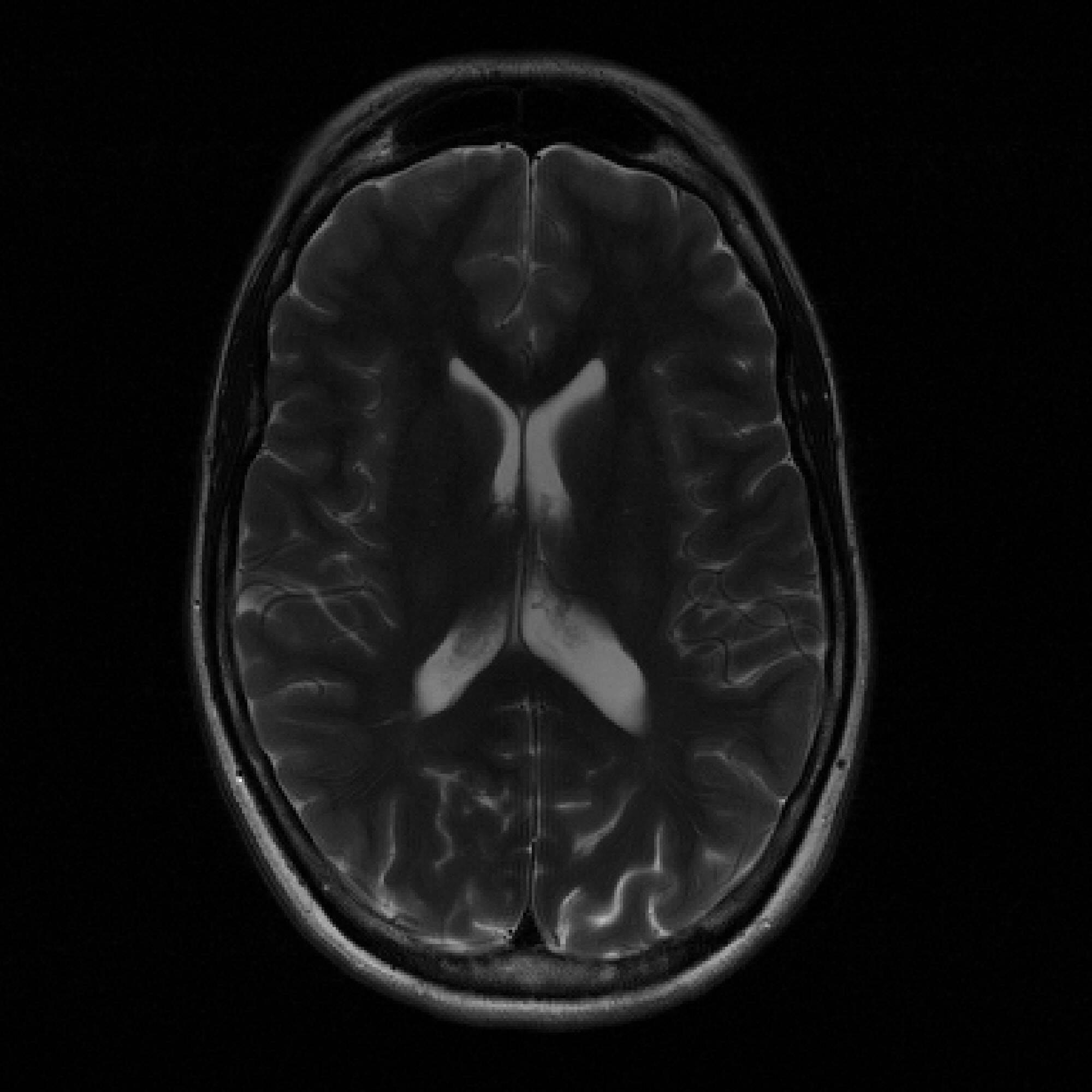} {0.3,0.45} {0.70,0.45} &
    \mrizoom{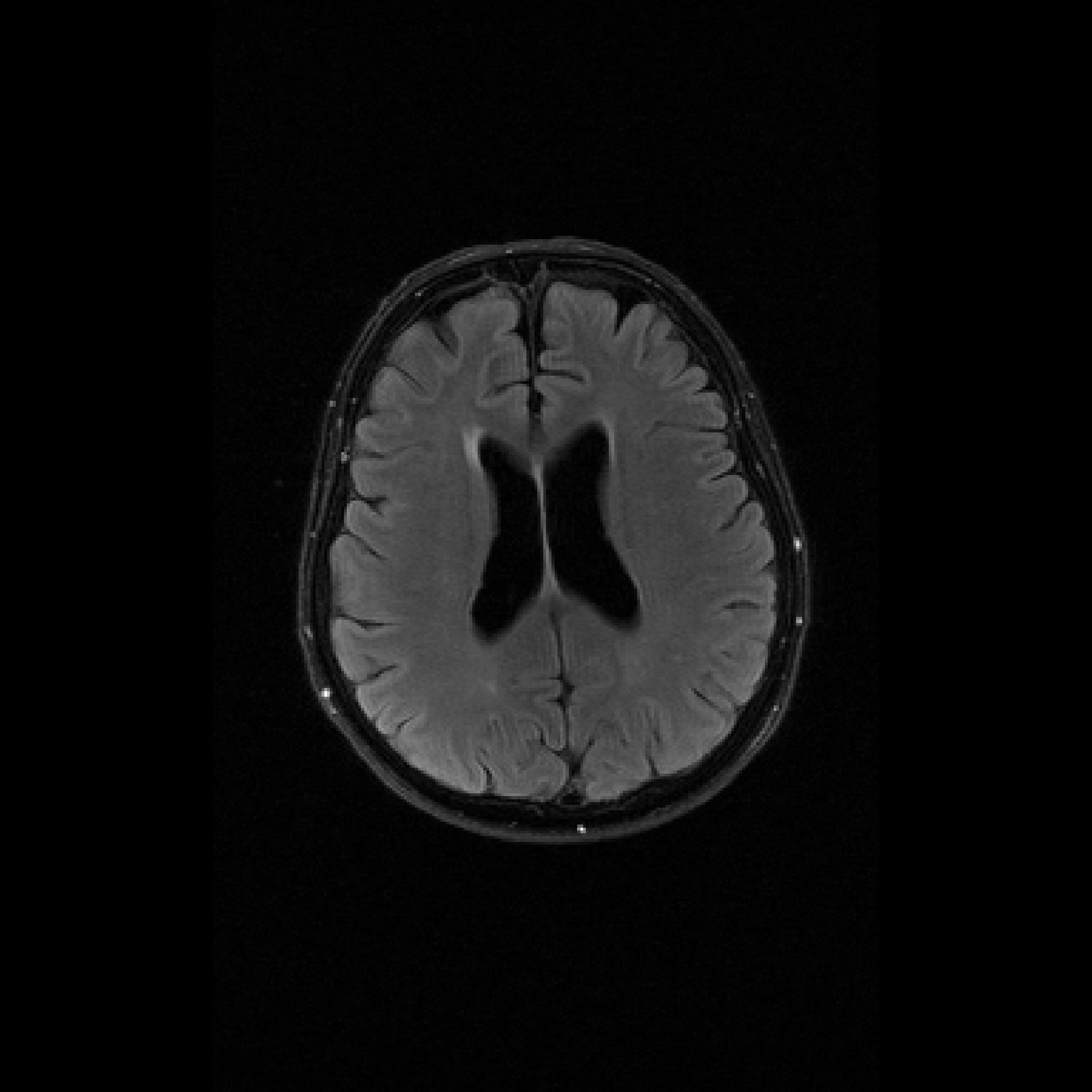}  {0.50,0.28} {0.50,0.69} &
    \mrizoom{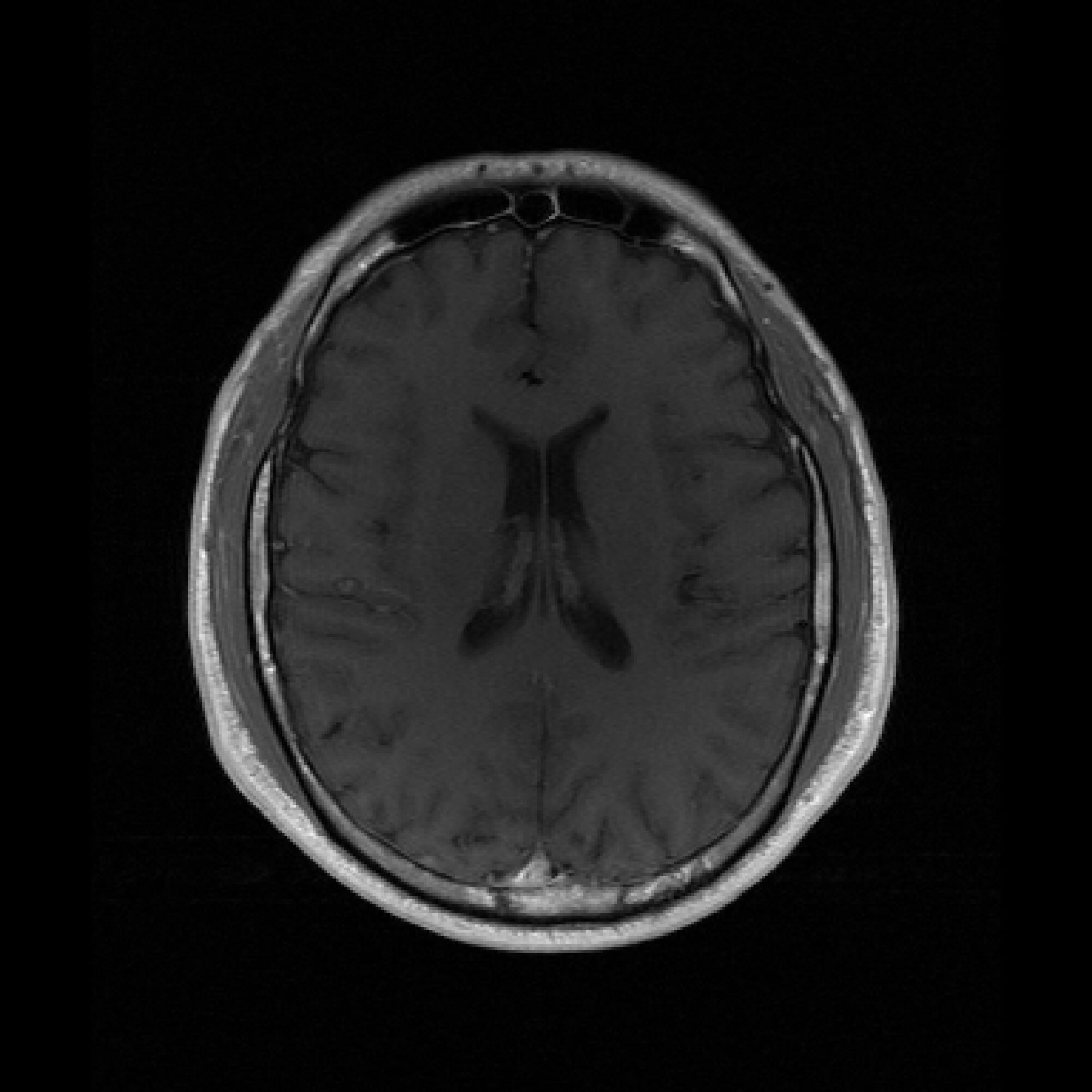}    {0.3,0.45} {0.65,0.45} \\
    \mrizoom{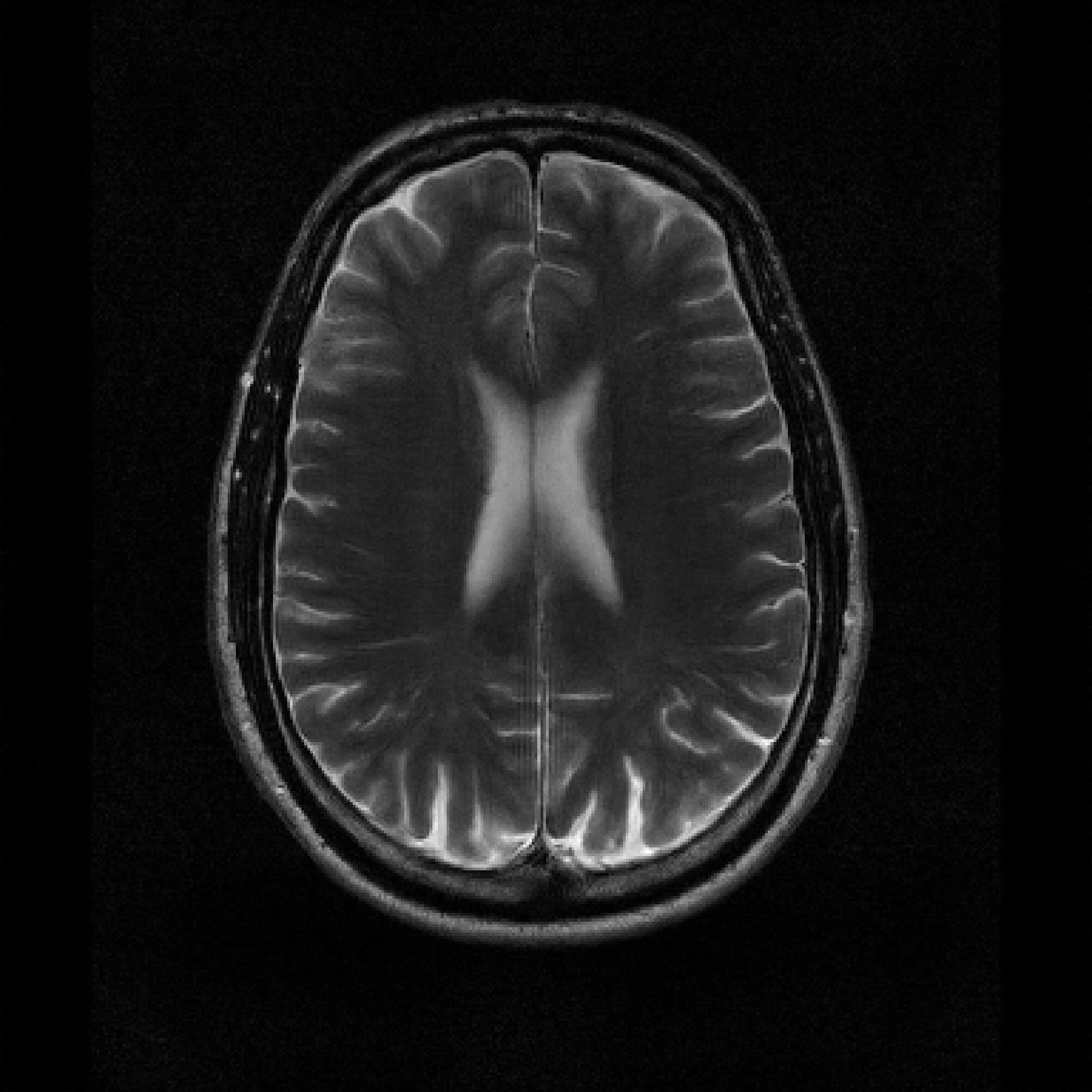}    {0.28,0.44} {0.47,0.25} &
    \mrizoom{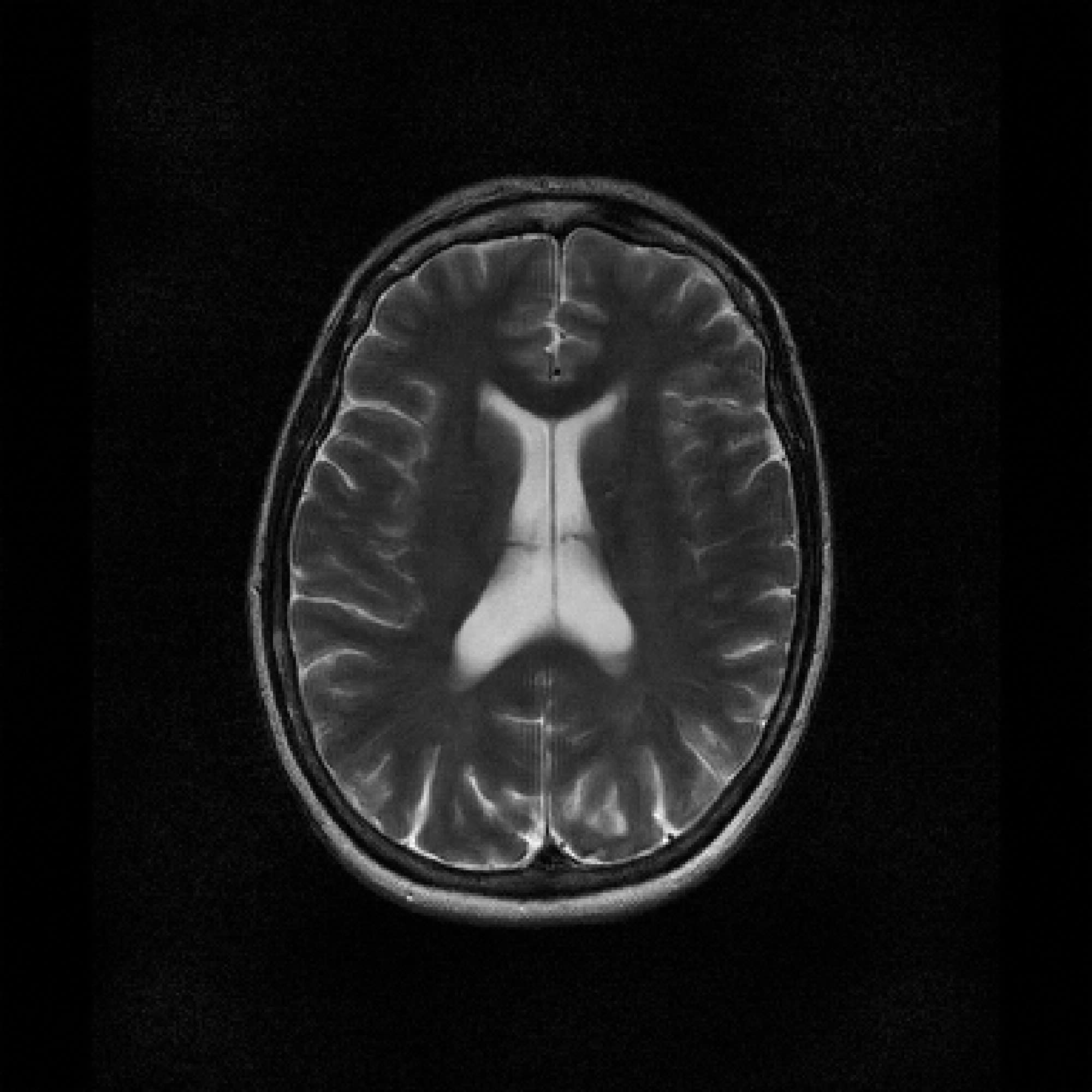}    {0.35,0.45} {0.5,0.25} &
    \mrizoom{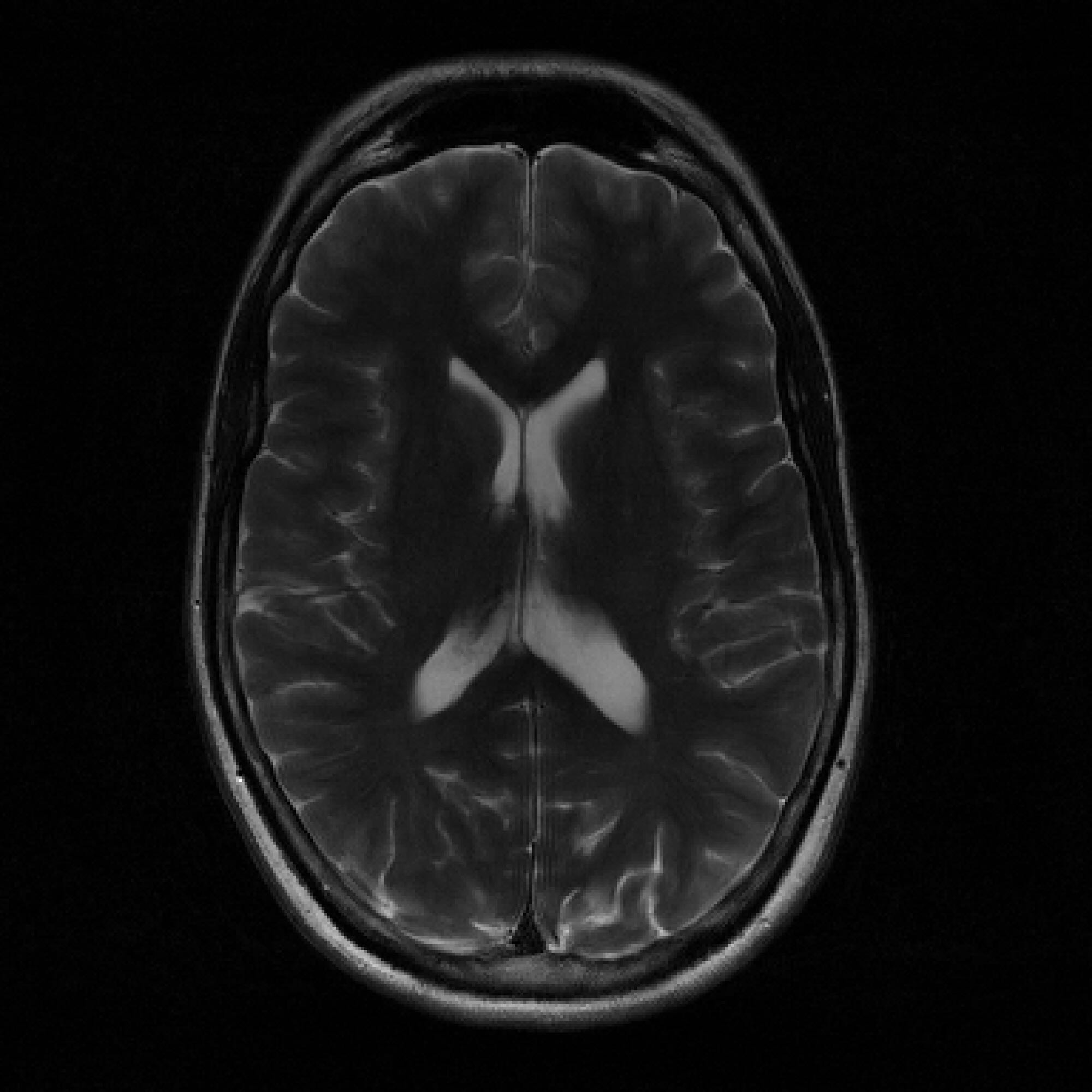} {0.3,0.45} {0.70,0.45} &
    \mrizoom{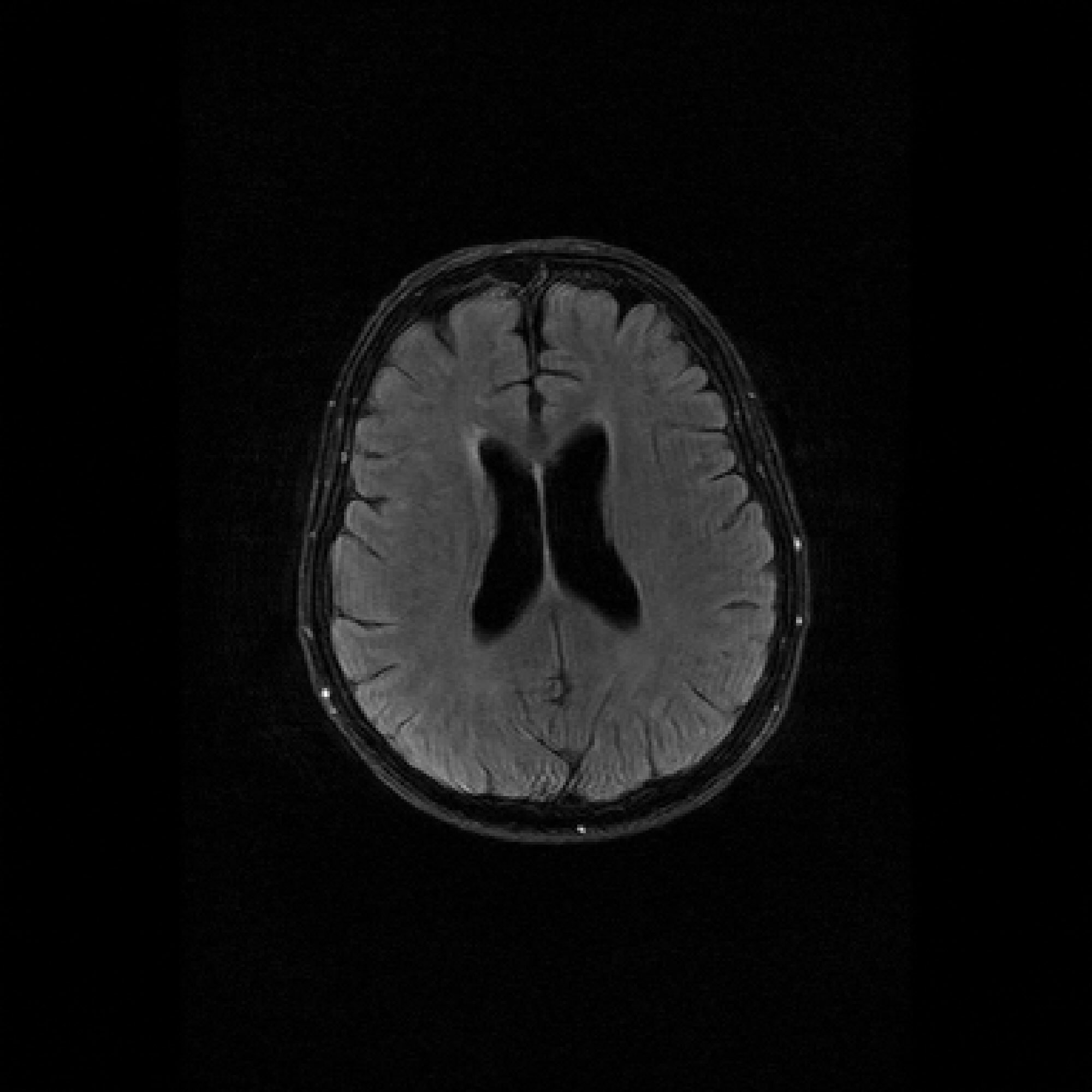}  {0.50,0.28} {0.50,0.69} &
    \mrizoom{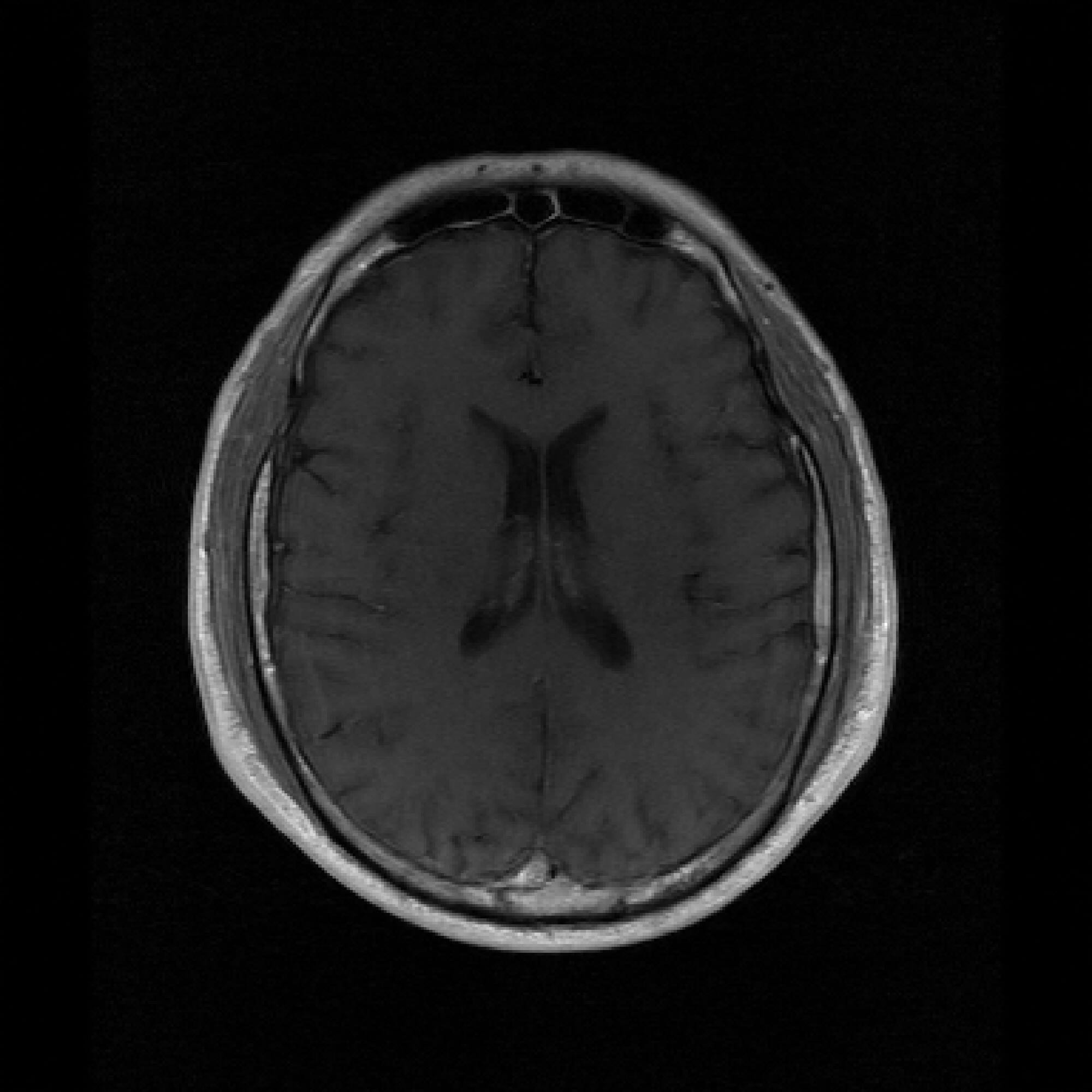}   {0.3,0.45} {0.65,0.45} \\
    \mrizoom{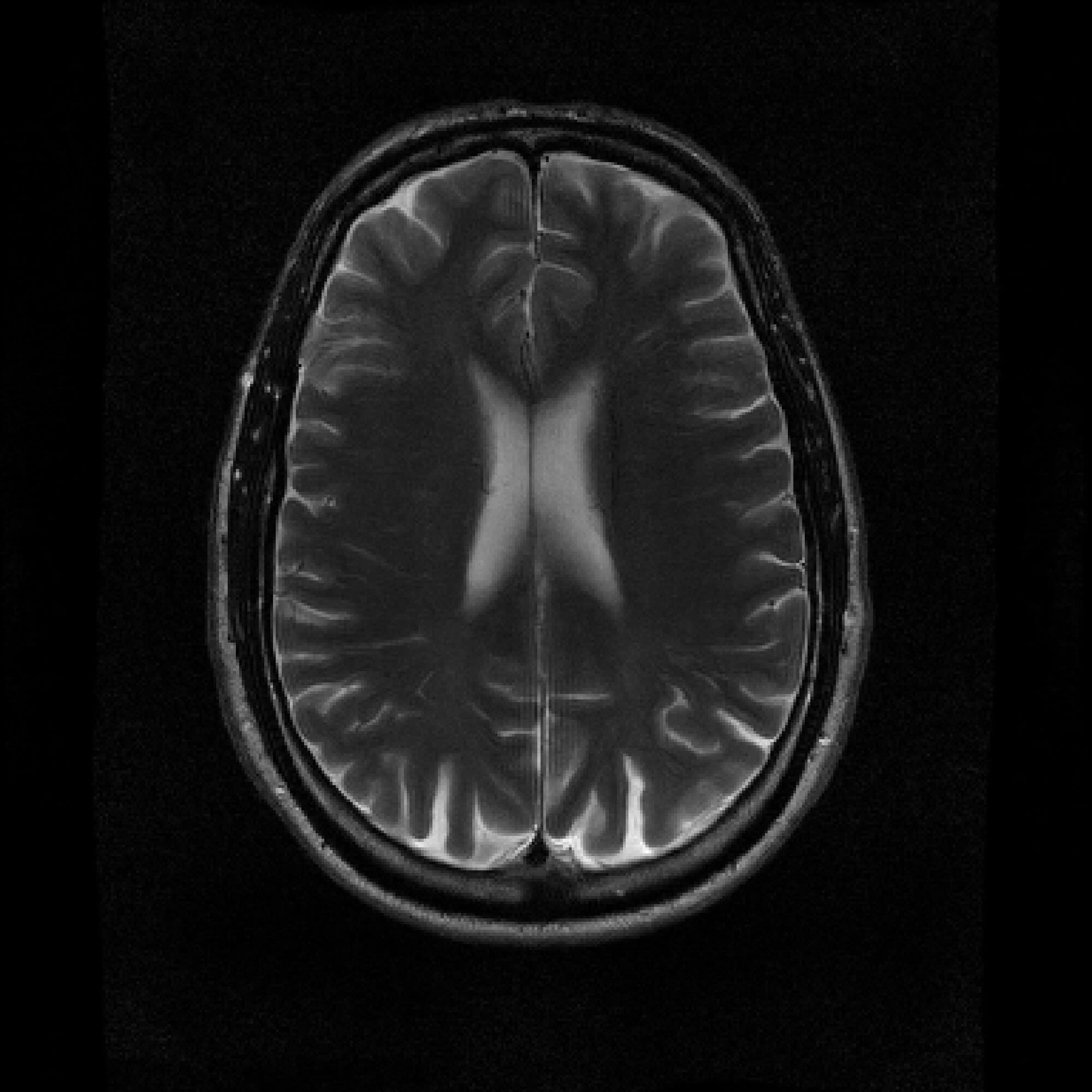}  {0.28,0.44} {0.47,0.25} &
    \mrizoom{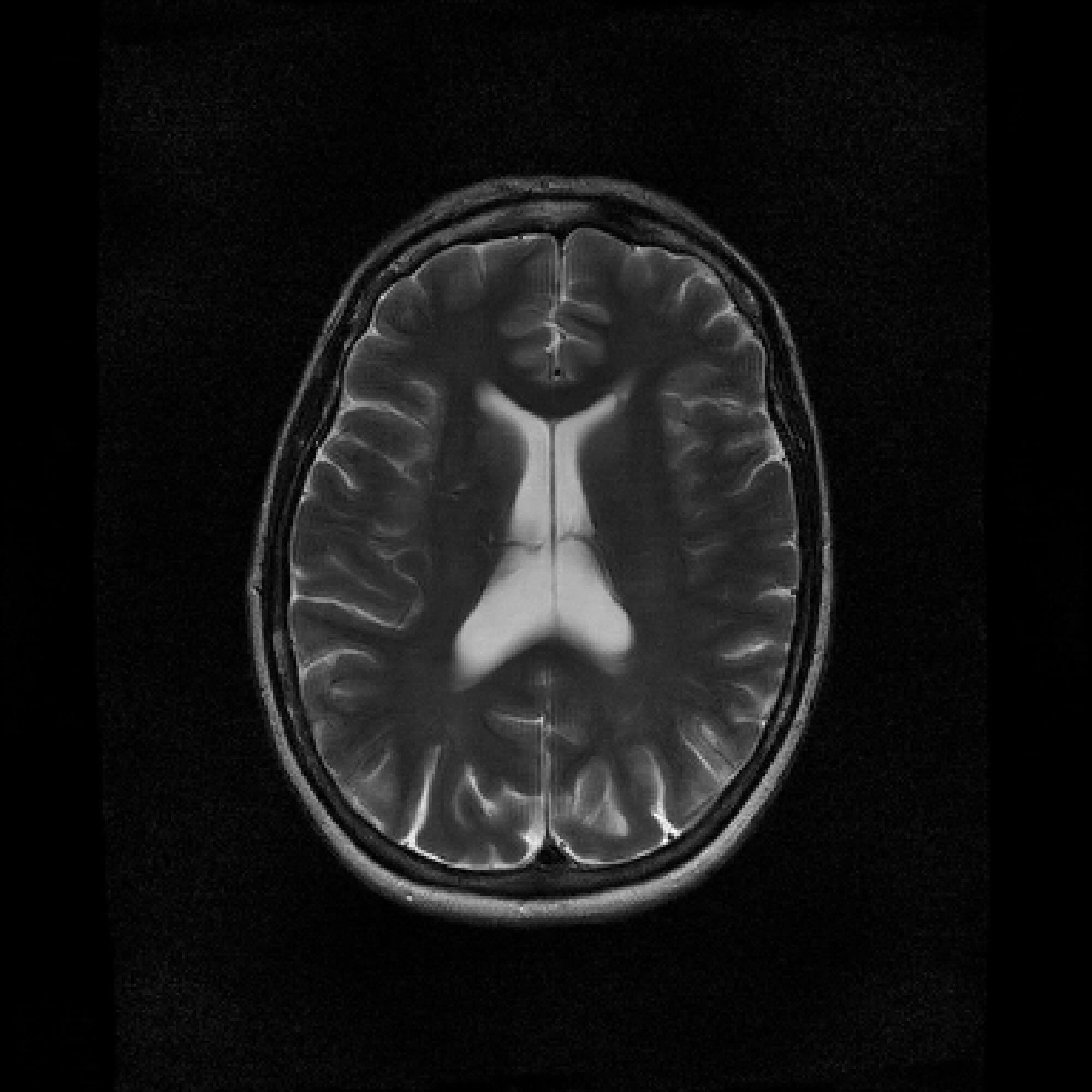}  {0.35,0.45} {0.5,0.25} &
    \mrizoom{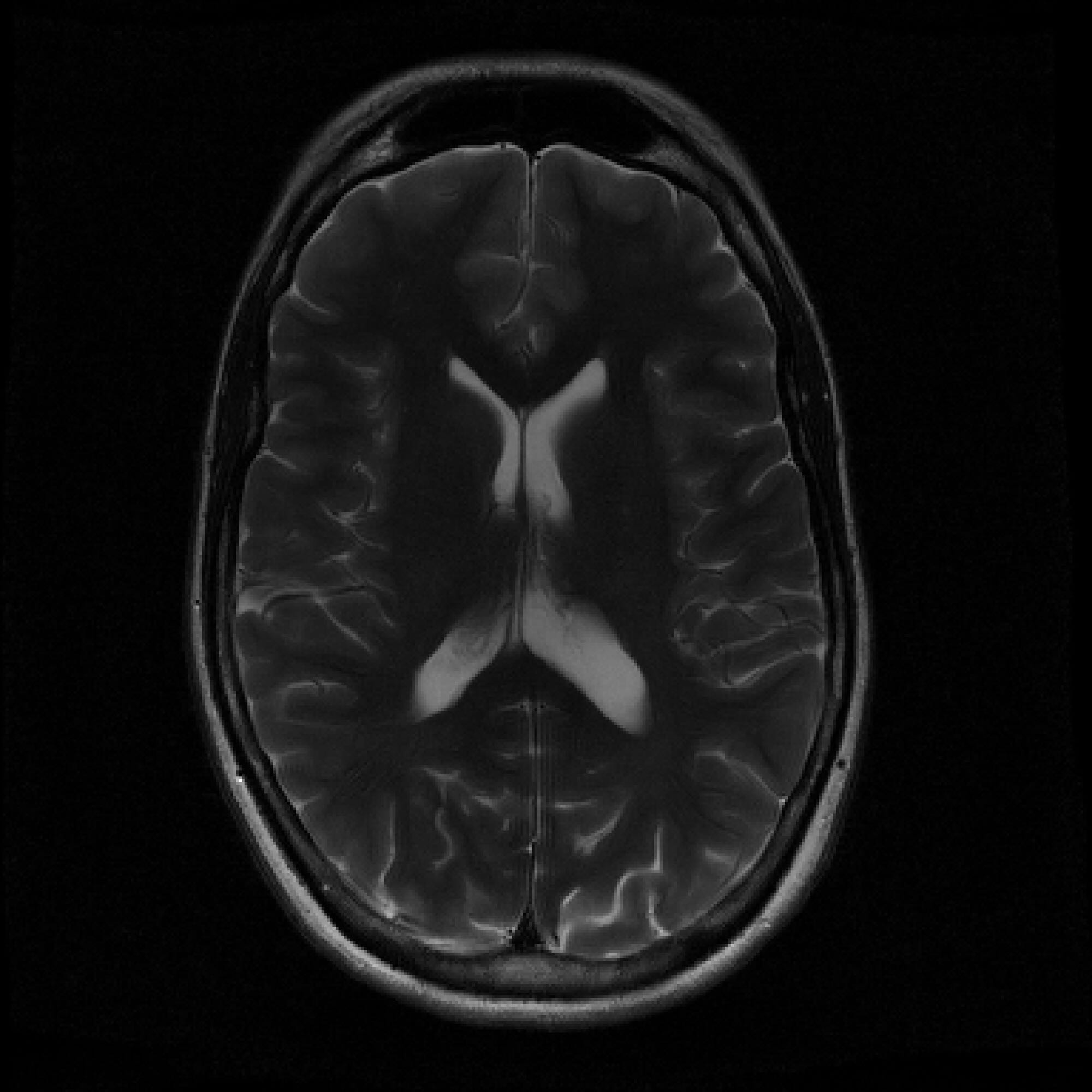} {0.3,0.45} {0.70,0.45} &
    \mrizoom{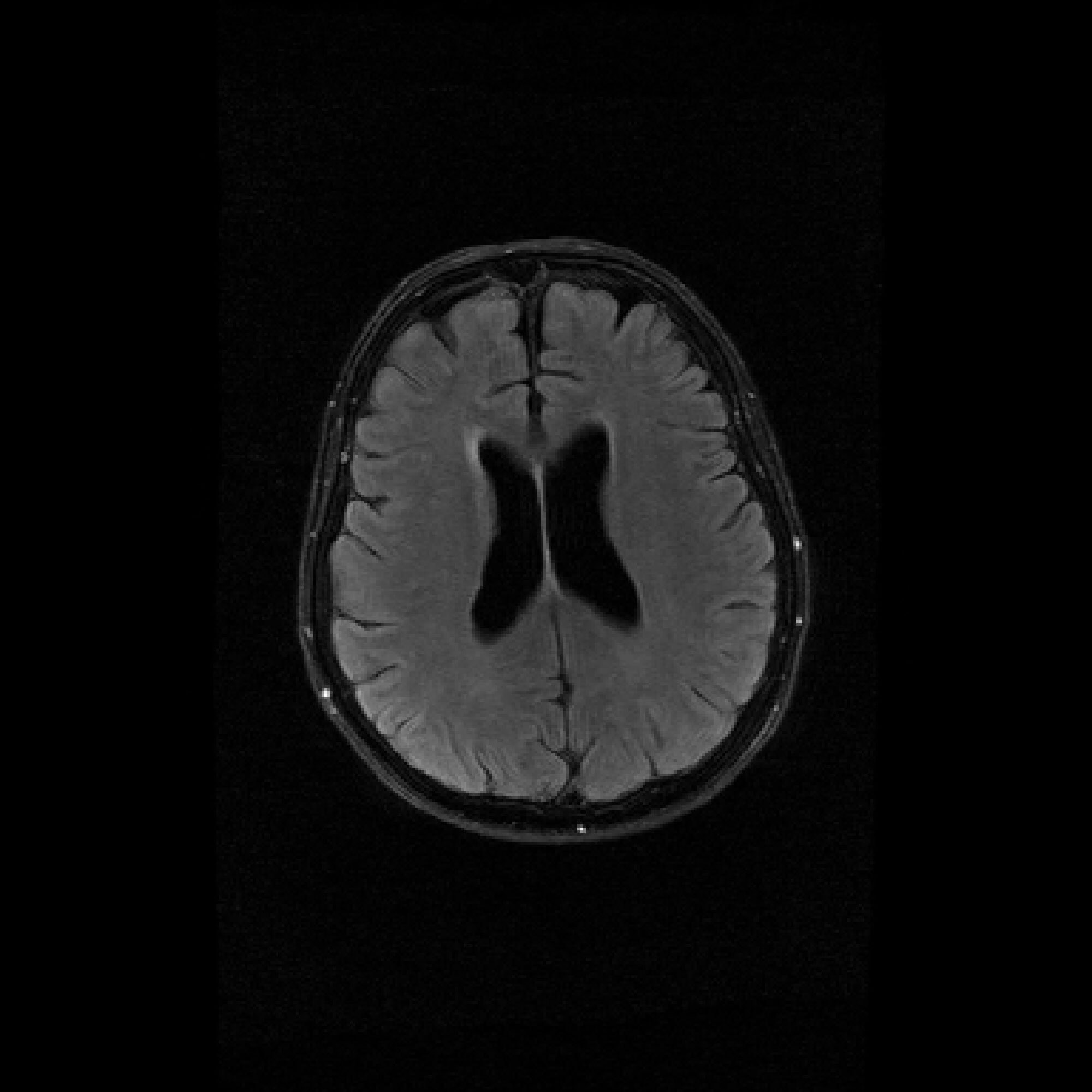}  {0.50,0.28} {0.50,0.69} &
    \mrizoom{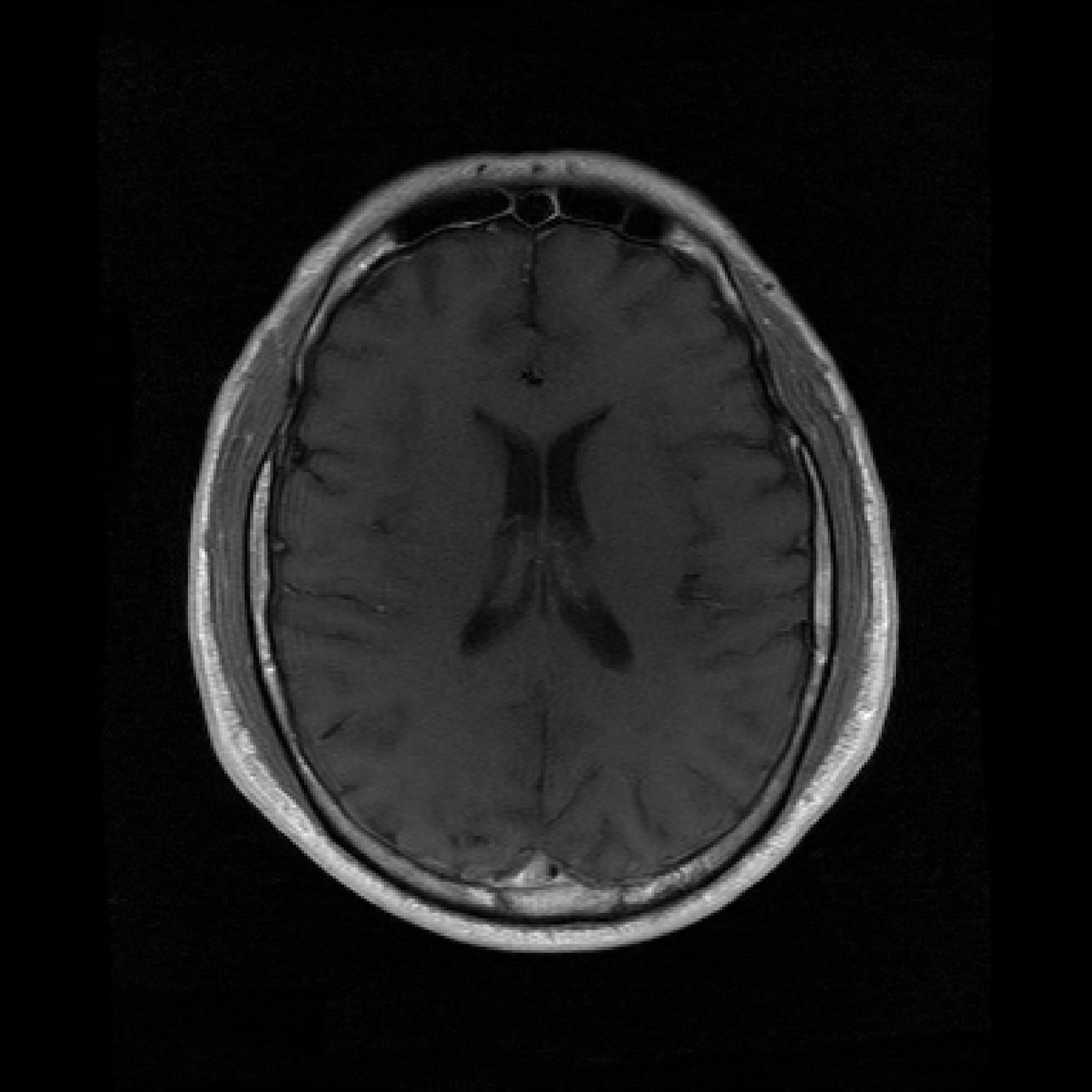}  {0.3,0.45} {0.65,0.45}
  \end{tabular}
  \caption{\textbf{Dataset Size $S=200$, Undersampling Rate $R = 7$ Inset Comparisons.} The top row shows the ground truth images. The middle row shows the FastMRI-EDM reconstructions. The bottom row shows the PaDIS-MRI reconstructions, which exhibit slight perceptual improvements at the insets.}
  \label{fig:mri_fullpage}
\end{figure*}

\end{document}